\RequirePackage{fix-cm}
\documentclass[smallextended]{svjour3}       
\usepackage[asymmetric]{geometry}
\smartqed  
\usepackage{graphicx}
\usepackage{amsfonts}
\usepackage{amsmath}
\usepackage{multirow}
\begin{document}
\title{Leveraging Multi-aspect Time-related Influence in Location Recommendation
}


\author{Saeid Hosseini\and Hongzhi Yin\and Xiaofang Zhou\and Shazia Sadiq}

\institute{
	Author's addresses: S. Hosseini, The University of Queensland, School of Information Technology and Electrical Engineering,
	QLD 4072, Australia; e-mail:ssaeidhosseini@gmail.com;H. Yin, The University of Queensland, School of Information Technology and Electrical Engineering,
	QLD 4072, Australia; e-mail:db.hongzhi@gmail.com; X. Zhou,
	S. Sadiq, The University of Queensland, School of Information Technology and Electrical Engineering,
	QLD 4072, Australia; e-mails: \{zxf, shazia\}@itee.uq.edu.au.
}

\date{Received: date / Accepted: date}

\maketitle

\begin{abstract}

Point-Of-Interest (POI) recommendation aims to mine a user's visiting history and find her/his potentially preferred places. 
Although location recommendation methods have been studied and improved pervasively, the challenges w.r.t employing various influences including temporal aspect still remain. Inspired by the fact that time includes numerous \textit{granular slots} (e.g. minute, hour, day, week and etc.), in this paper, we define a new problem to perform recommendation through exploiting all diversified temporal factors. In particular, we argue that most existing methods only focus on a limited number of time-related features and neglect others. Furthermore, considering a specific granularity (e.g. time of a day) in recommendation cannot always apply to each user or each dataset.\\
 To address the challenges, we propose a probabilistic generative model, named after \textit{Multi-aspect Time-related Influence} (MATI) to promote POI recommendation. We also develop a novel optimization algorithm based on \textit{Expectation Maximization} (EM). Our MATI model firstly detects a user's temporal multivariate orientation using her check-in log in Location-based Social Networks(LBSNs). It then performs recommendation using temporal correlations between the user and proposed locations. Our method is adaptable to various types of recommendation systems and can work efficiently in multiple time-scales. Extensive experimental results on two large-scale LBSN datasets verify the effectiveness of our method over other competitors.\\ \\
 Categories and Subject Descriptors: H.3.3 [\textbf{Information Search and Retrieval}]: Information filtering; H.2.8 [\textbf{Database
 Applications}]: Data mining; J.4 [\textbf{Computer Applications}]: Social and Behavior Sciences 
 
\keywords{Multi-aspect Time-related Influence \and Hybrid Temporal Location Recommendation \and Location-based Service}
\end{abstract}

\clearpage 
\section{Introduction}
\label{introduction}
With the ubiquity of GPS-enabled smartphones, using Location-based Social Networks(LBSNs) has become an essential part of the daily life. People can easily socialize and share their check-in data through such mediums (e.g. \textit{Foursquare, Yelp, Gowalla, Loopt and Google places}). Specifically, when an individual presses the check-in button, she reports her location and further information such as time stamp and artifacts including textual contents, photos, videos and so on(\cite{Zheng2011},\cite{Symeonidis2014}). On the one hand, LBSN users share such valuable spatio-temporal information generously. On the other hand, by mining the rich user check-in data, Point-Of-Interest(POI) recommendation systems assist the users in exploring new attractive venues which affirm numerous benefits for all stakeholders in the LBSN ecosystem.\\
A check-in in LBSNs is interpreted as a digit entry in the user-location matrix where it can indicate the frequency of a user's visit to a location \cite{Cheng2012} or a binary value (e.g. \cite{Ye2011,Yuan2013}) to denote whether she has visited the location or not. In a positive-only manner, missing an entry signifies that we don't know whether the user doesn't like the venue or is basically unaware of it. Nevertheless, the primary problem regarding location recommendation services is to suggest a list of new interesting POIs to the query user. In light of excessive \textit{sparsity} (also called as \textit{scarcity}) observed in user-location matrices (\cite{Wang2015}), recommending POIs to users in LBSNs is extremely challenging.\\
The challenge has been tackled through various methods such as traditional models (\cite{Tong2006,Xiang2010}) which predict user's interest on POIs through \textit{Random Walk and Restart} and dominant \textit{Collaborative Filtering (CF)} (\cite{Ye2011,Levandoski2012,Yuan2013,Yin2015}) that infer the user's preference regarding each unvisited POI which are categorized into memory-based and model-based methods (\cite{Cheng2013a}). In reality, numerous methods (\cite{Yin2016,Gao2013,Levandoski2012,Ye2011}) provoke the problem through alleviating sparsity in user-location matrix. In addition, an extensive amount of research has been dedicated recently to take various influences into consideration to enhance recommendation process. \textit{Geographical} (\cite{Ye2011,Yuan2013,Liu2013}), \textit{social} (\cite{Cheng2012,Gao2012}), \textit{context-oriented} (\cite{Yin2013,Yin2015}) (e.g. text contents and word-of-mouth) and \textit{temporal} effects are among commonly utilized factors \cite{Yin2015}. Despite the vital beneficial role of temporal influence (\cite{Zhao2016a}), not an adequate amount of research has been devoted to deal with the time-aware location recommendation. As a matter of fact, regular User-Time-POI hexahedron (UTP) is more scattered compared to the User-Location matrix. While we have an insufficient number of records regarding a user's check-in at a particular location, predicting the time of the visit seems more problematic. Anyhow, we consider time-related information to enhance the effectiveness of location recommendation systems.\\
Thinking about time, POI recommenders have so far employed three temporal attributes of periodicity, consecutiveness, and non-uniformness(\cite{Cheng2013a,Cho2011,Gao2013,Yuan2013,Zhao2016,Fang2016}). Periodicity (\cite{Cho2011}) states that a user's movement in different locations has an approximate periodical replication. For example, a typical user would mostly perform check-ins near her workplace throughout the day and at her own property during after hours. Consecutiveness or Successive attribute (\cite{Zhang2015b,Cheng2013a}) claims that there are certain locations which are visited in a sequential order during a limited time constraint. Finally, non-uniformness declares that the check-in behavior of LBSN users vary in different temporal periods (i.e. one's activity pattern is work-oriented during weekdays and related to entertainment throughout the weekends) (\cite{Gao2013}). Inspired by the fact that the time dimension comprises numerous \textit{granular slots} (e.g. minutes, quarters, hours, days and etc.), and some are subset of others, we propose the fourth attribute named as \textit{Temporal Subset Property} (TSP).\\
Some other research also reveals that the time factor can be treated either discretized (\cite{Yuan2013,Gao2013,Zhao2016,Fang2016,Yin2016,Deveaud2015}) or continuous (\cite{Zhang,Yin2015}). Those using the time in a continuous manner claim that choosing the proper time interval is not feasible (\cite{Yin2015}). However, discrete-time is the base of our daily lives. We set our appointments, meetings, and events using predefined time slots. Additionally, urban arrangements are planned by discretized values (e.g. a sample supermarket chains in Australia close at 6pm except on Thursdays in which they serve the customers until 9pm), hence it makes sense for us to use discrete-time in our paper. However, previous work \cite{Zhao2016,Yin2016,Fang2016,Zhang2015b,Deveaud2015,Gao2013,Yuan2013,Zhang} that integrated discretized temporal information such as the hour of the day or day of the week cycles into POI recommendation considered only a single or two temporal granularities to avoid complexity and overfitting issues \cite{Zhang2015b}. In reality, time is multi-aspect and user check-in behaviors are simultaneously influenced by multiple temporal effects with different cycles or granularities. Also rather than configuring the method (e.g. \cite{Gao2013}) to work under specific time-related intervals, it is better to devise a solution which can include multiple temporal factors to promote recommendation systems.\\
Accordingly, our observation on two public LBSN data sets (Section \ref{Data_Set}) shows more than 40\% of locations, explored by at least 8 users, are mostly visited in their popular times (e.g. a bar is mostly visited during after hours). Hence, a location $l_j$'s probability to be visited by a user $u_i$ increases when $u_i$ owns prior check-ins during the times when $l_j$ is visited more. As the time of a visit can be declared through several dimensions (minute of hour, hour of the day, the day of the week and so on), we can conclude that LBSN users and locations correlate with each other temporally in a multi-aspect way. To summarize, what we are seeking, throughout this paper, is ``\textit{What kind of model do we have to choose to comprise all temporal dimensions in POI recommendation? How to mitigate sparsity in a hierarchical set of UTP matrices where each one is associated with a single temporal dimension? Finally, how can we use this perception to enhance POI recommendation systems?}''\\
To this end, we initially aim to reduce sparsity. We select an optimum number of users via non-replacement stratified sampling model (\cite{Liberty2016}). Subsequently, we extract a list of user-time-POI matrices for every sampled user. Each of these cubes is associated with a temporal dimension. We then apply similarity metrics to find homogeneous parts in each dimension (e.g. similar hours, days and etc). Consequently, through aggregating the evidence captured from every user, we can reach the final similarity maps for each temporal granularity. We also utilize matrix factorization to compute missing values in similarity maps (e.g. we may not have enough evidence to find the similarity between 2am and 3am in the hour of a day dimension). Subsequently, we use the Bottom-up Hierarchical Agglomerative Clustering (\textit{HAC}) \cite{Das2014} to merge similar partitions in various granularities. This constructs primary multi-aspect \textit{Temporal Slabs}. Each temporal slab includes a set of similarly merged parts from various scales. Such preprocessing mitigates sparsity involved in UTP matrices as the user's check-in at a given time can be estimated by the check-ins throughout similar times.\\
Moreover, we propose a probabilistic generative model, named after \textit{Multi-aspect Time-related Influence} (MATI) which consumes constructed temporal slabs to recommend a temporally correlated list of new POIs to the query user. As each user's location history is insufficient, we utilize a novel Expectation Maximization algorithm to infer latent parameters and subsequently compute both the depth and extent of temporal similarity between the query user and each of proposed locations. While the depth of correlation is computed through aggregation of the joint probabilities of the user, location, and all latent temporal factors, the extent of the correlation is calculated via Jaccard coefficiency among temporal slabs associated with the query user and the proposing location. We theoretically prove that the model can simultaneously integrate multiple latent temporal impacts in the recommendation task. Nonetheless, not necessarily all the users would follow leveraged temporal patterns. For instance, owing to holidays, a user may go to a restaurant on Monday at 10am and go to a bar successively. While, such a behavior has the least likelihood for the majority of other users who are at work. Therefore, in a hybrid framework, we firstly detect whether each query user is adequately affected by the time factor or not. On the other hand, our method mimics how the user and the set of top N proposed locations share a commonly acceptable check-in behavior. If the computed metric is fallen in the well-tuned range, the multivariate temporal influence will be implanted, and vice versa.\\
To summarize, this paper focuses on the problem of enhancing location recommendation task in social networks. We have previously presented our study regarding the effects of a single temporal granularity in Hosseini et al. \cite{Hosseini2016}. This article extends Hosseini et al. \cite{Hosseini2016} through utilizing various dimensions of temporal influence and in-depth performance analysis. Specifically, this article provides the following new contributions: first, our previous model addresses the role of a single temporal effect in recommendation, while in this work, our proposed method takes an unlimited number of concurrent dimensions into consideration; second, we propose an optimized approach to retrieve multi-aspect similarity maps which also mitigates sparsity; third, MATI as a latent generative model can predict user's time oriented mobility patterns. Moreover, The ultimate advantage of our model in capturing all temporal aspects can enhance various user-item based recommendation systems disregarding the level of density; We also review temporal trends of location recommendation systems in related work. To the best of our knowledge, no prior work attempted to take into account both multi-aspect(MATI) and subset(TSP) temporal features to enhance location recommendation.\\
To sum up, the main contributions of this paper are listed as follows:
	\begin{itemize}
        \item Our method exploits \textit{multi-aspect temporal slabs} through merging similar temporal slots in various scales\footnote{A temporal slot, scale, and dimension (e.g. Hour, Day and etc.) are used interchangeably in this paper unless noted otherwise.}. The size of the dataset can be huge and the users' check-ins during certain slots might be low in number. Therefore, in a novel procedure, our method estimates proper slabs through processing of a minimum subset of the dataset by employing both stratified sampling and matrix factorization.
        \item Our generative model (MATI) can incorporate as many latent temporal granularities as required where each of them will represent a temporal scale. It enhances the results of the recommender framework through leveraging the multi-aspect temporal correlation between LBSN users and POIs. The model reflects the extent of the shared temporal activity as well as the depth of time-related visibility patterns between each pair of sample user $u_i$ and location $l_j$.
        \item We also devise an automatic hybrid solution which can decide whether each user can benefit from multivariate temporal influence or not.
	\end{itemize}
    The rest of this paper is structured as follows. We start with a review of our prior work (\cite{Hosseini2016}) in section \ref{Temporal_Influence} which exploits a univariate time-related influence. We then continue with the extension which considers the multivariate aspect of the time using a comprehensive probabilistic approach devised based on temporal latent factors. Subsequently, section \ref{evaluation} provides experimental results. Related research work is surveyed in section \ref{Related_Work}. We finally close this paper in section \ref{conclusions} which offers promised future directions and conclusive remarks.
	\section{Single Slot Temporal Influence}
	\label{Temporal_Influence}	
\begin{figure}[!htp]	
		\centering
		\tiny
		\minipage{0.40\textwidth}
		\centering
		\includegraphics[width=\linewidth]{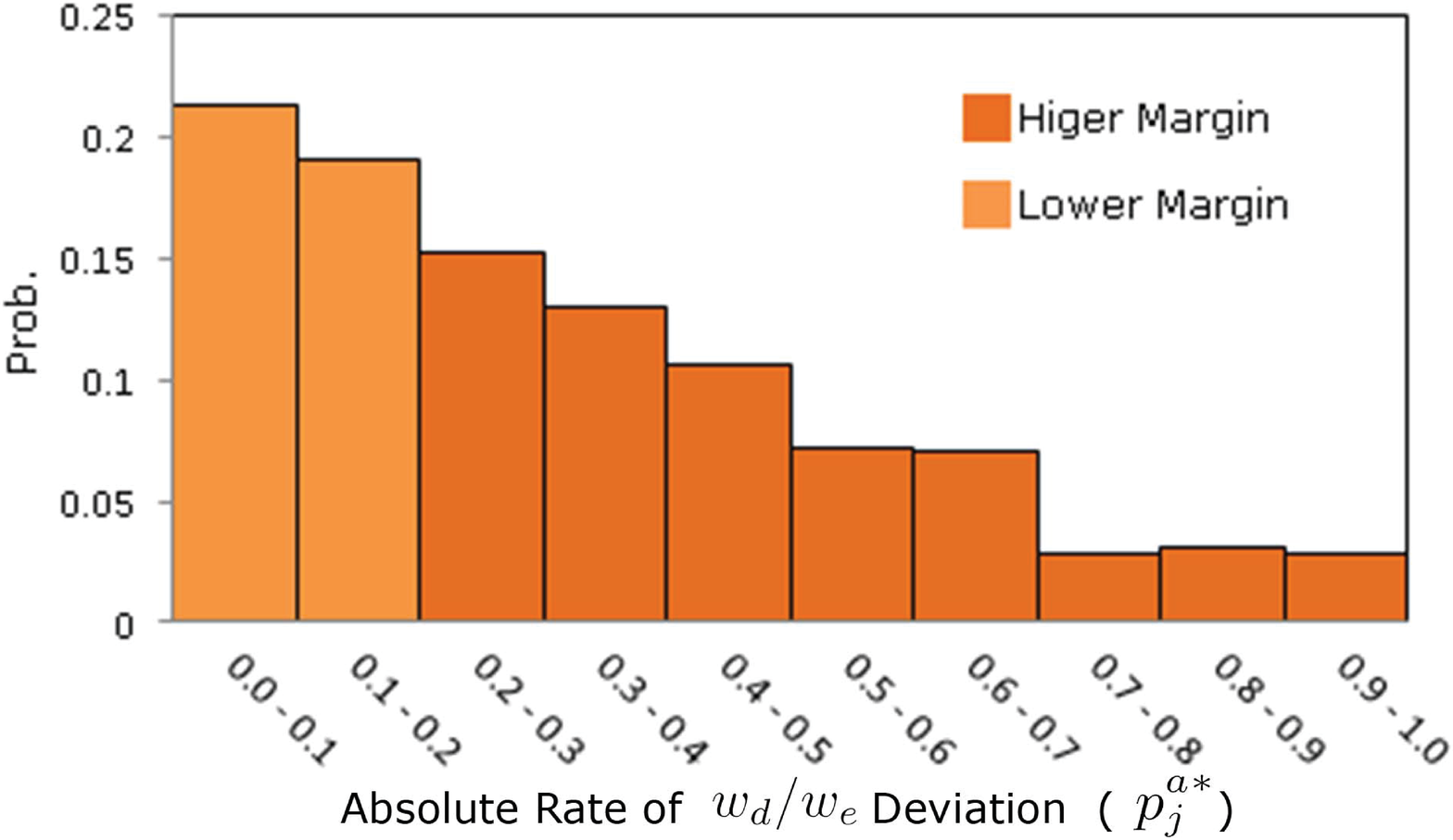} 
		\small (a) Foursquare
		\endminipage\hfill
		\minipage{0.40\textwidth}
		\centering
		\includegraphics[width=\linewidth]{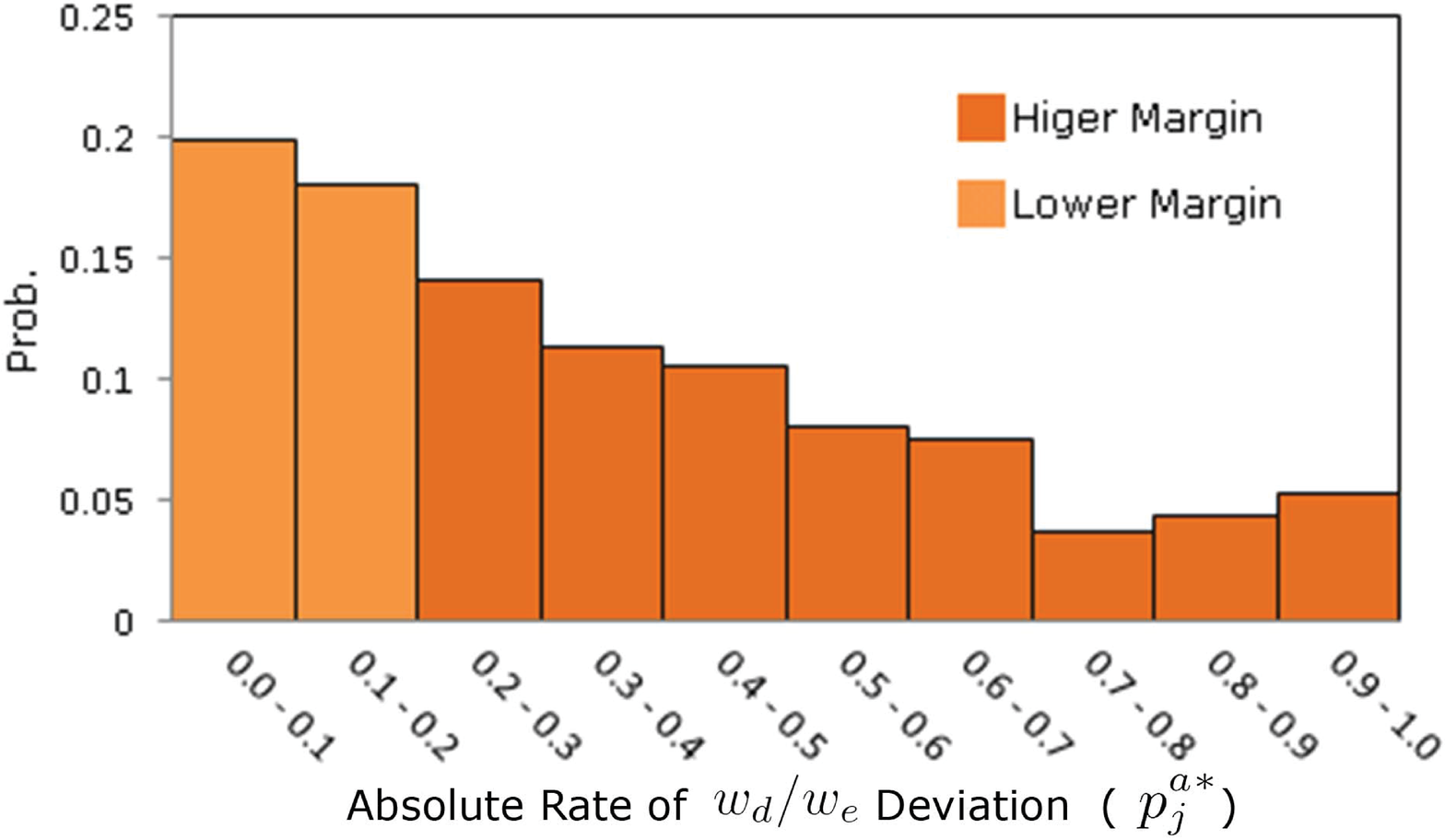} 
		\small (b) Brightkite
		\endminipage
		\caption{\small Observation on Absolute POI Act}
		\label{POI_Act_Observation}
	\end{figure}

\begin{figure}[!htb]	
		\centering
		\minipage{0.40\textwidth}
		\centering
		\includegraphics[width=\linewidth]{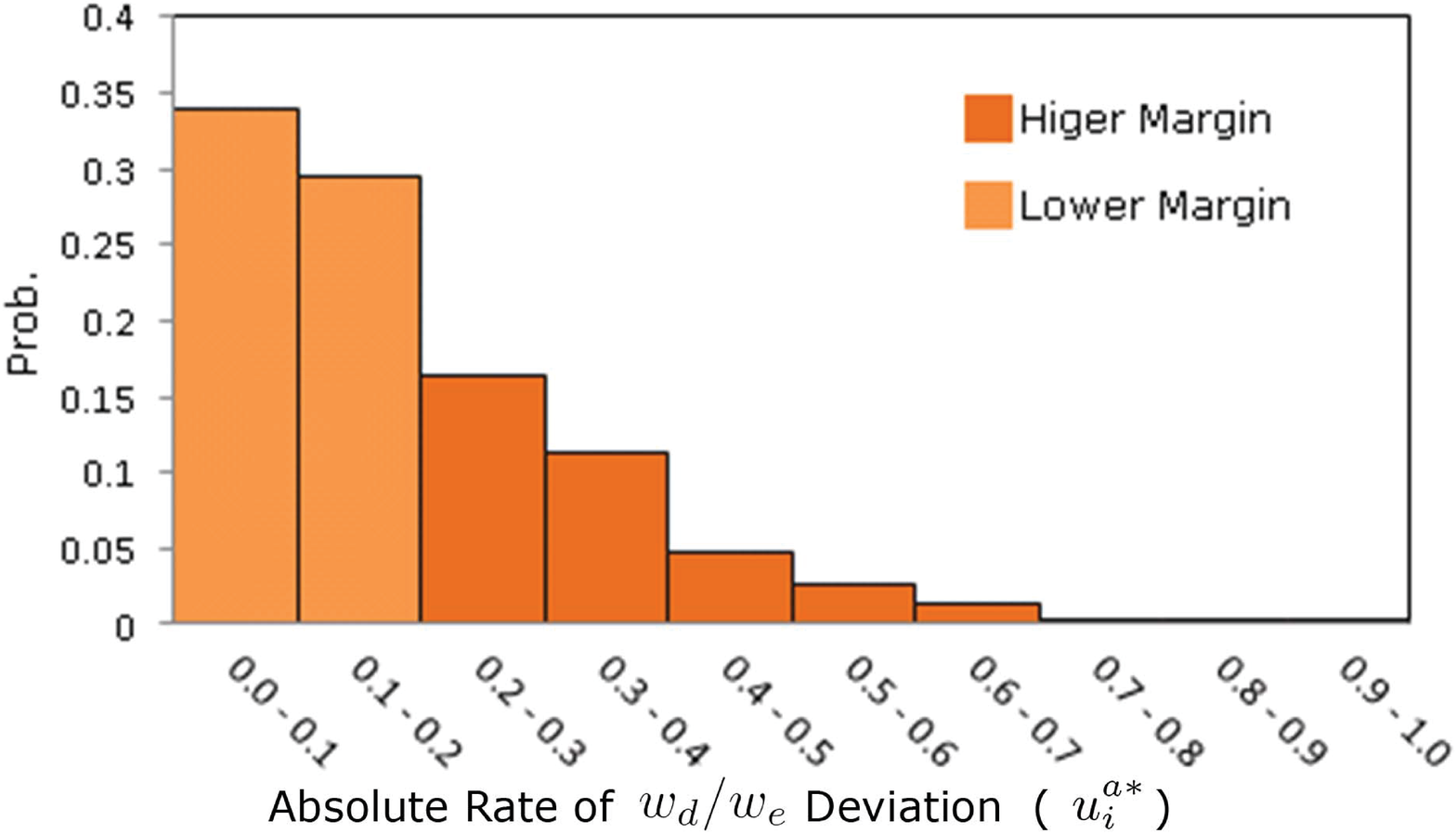} 
		\small (a) Foursquare
		\endminipage\hfill
		\minipage{0.40\textwidth}
		\centering
		\includegraphics[width=\linewidth]{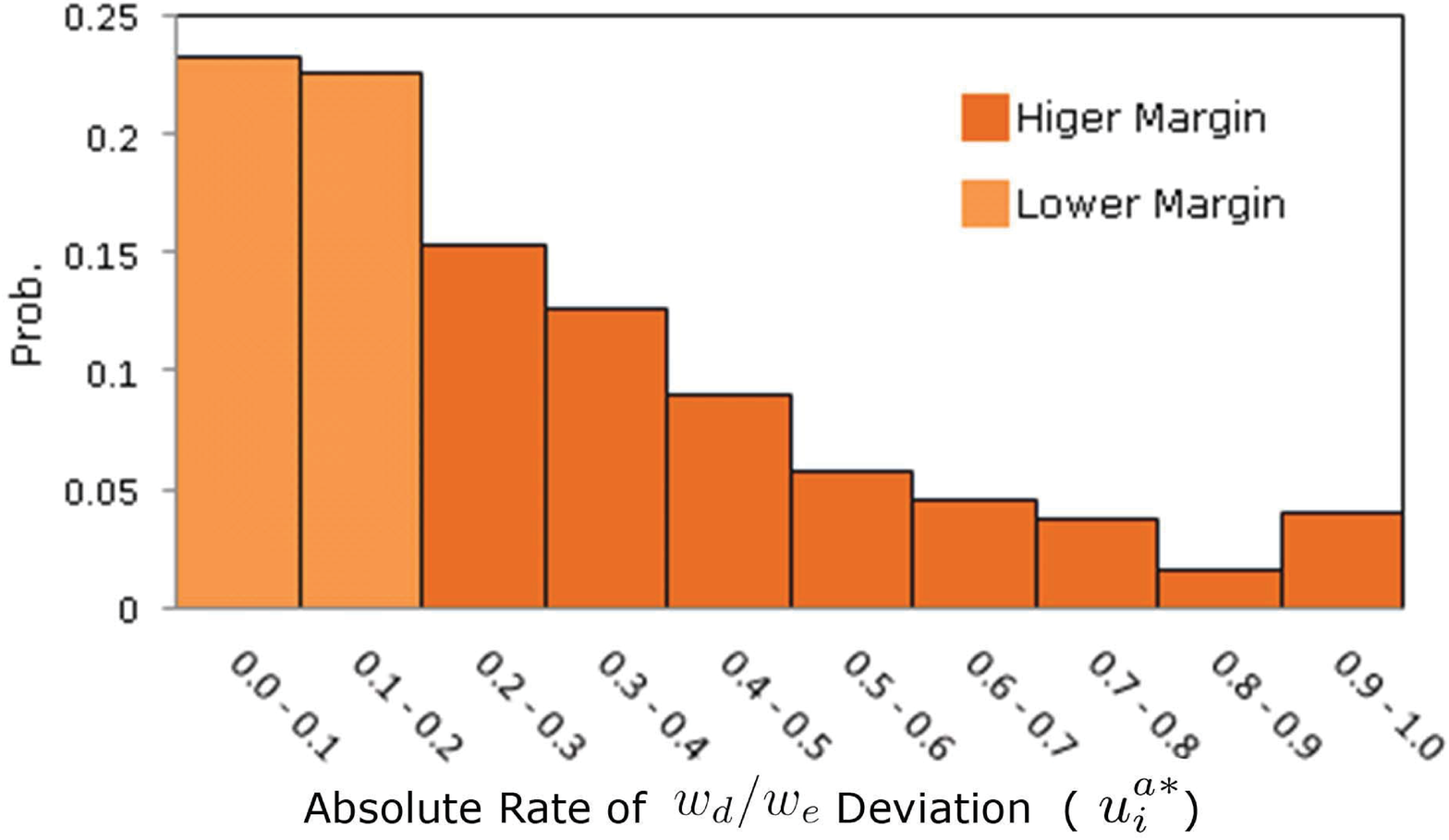} 
		\small (b) Brightkite
		\endminipage
		\caption{Observation on Absolute User Act}
		\label{UserAct_Observation}
	\end{figure}

	Intuitively, selecting a specific temporal granularity (e.g. hour of the day) and leave others (e.g. minute of hours, the day of the week and ...) unattended is not legitimate. In addition, there is not a clear reason why one scale can be preferred versus others. Such an argument justifies the necessity of a model which can integrate multiple temporal scales into the location recommendation methods. For simplicity, we begin with a \textit{review of our previous work} (\cite{Hosseini2016}) in this section. This merely considers a single temporal aspect. Nevertheless, it can help gaining a primary understanding about the visibility correlations of user-location pairs.\\
	From a univariate temporal perspective, as visited locations during weekday and weekend are substantially different, we can study weekly intervals to see how it can promote the effectiveness of POI recommendation system. People usually visit entertainment venues during weekends and work related places throughout weekdays. Hence, we can develop a new method to perform recommendation, based
	on temporal weekly alignments of users and POIs. In our prior work \cite{Hosseini2016}, we developed a probabilistic model which detects a user's temporal orientation based on visibility weights of POIs visited by her during weekday/weekend cycles. Consequently, the system proposes locations based on her interest toward either of weekdays or weekends.\\
	To this end, we firstly set up two observations based on primary definitions to verify that certain POIs and users are aligned toward either weekday or weekend.		
	\begin{definition}
		\label{def:POI_ACT}
		\textbf{(POI Act)} Given a set of POIs $\mathbb{P}=\{p_1, p_2,\dots, p_n\}$, each $p_j$ ($\forall p_j \in \mathbb{P}$) has a POI Act denoted as $p_j^a$ (Eq. \ref{eq:POI_ACT}), which is the margin value ($[ -1,1 ]$) between its probabilities to be visited during weekday ($w_d$) and weekend ($w_e$).
	\end{definition}
	\begin{equation} 
	\small
	\label{eq:POI_ACT}
	p_j^a=\frac{W_j^d}{N_j}-\frac{W_j^e}{N_j}
	\end{equation}
	Here, $W_j^d$ and $W_j^e$ denote the number of visits at $p_j$ during $w_d$ and $w_e$. Also, $N_j$ is its total number of visits. If $p_j^a$ is greater than zero, it exhibits an alignment toward $w_d$; if it is less than zero, it shows that $p_j$ is visited more during $w_e$; otherwise (if $p_j^a=0$), $p_j$ will be neutral (not temporally aligned)\\	
	\begin{definition}
		\label{def:USER_ACT}
		\textbf{(User Act)} Given a set of users $\mathbb{U}=\{u_1, u_2,\dots,u_n\}$, we define that each $u_i$ ($\forall u_i \in \mathbb{U}$) has a User Act denoted as $u_i^a$ (Eq. \ref{eq:USER_ACT}) which is the margin value ([-1,+1]) between probabilities of her $w_d$ and $w_e$ visits.
	\end{definition}
	\begin{equation}
	\small
	\label{eq:USER_ACT}
	u_i^a=Avg_i^d-Avg_i^e
	\end{equation}
	$Avg_i^d$ and $Avg_i^e$ are probabilities for $u_i$ to visit locations during $w_d$ and $w_e$ respectively. If $u_i^a$ is greater than 0, it reflects $u_i$'s temporal preference toward $w_d$ and if it is less than 0, it indicates that she is more interested in $w_e$.\\
	Subsequently, we continue with observations to perceive the single aspect of weekday/weekend cycles in LBSN users' behavior. As certain POIs and users can be oriented toward $w_d$ or $w_e$, we can use threshold $T$ to reflect the extent of alignment. $T$ is set to $\frac{1}{7} \approx 15\%$ which follows the uniform distribution of locations for each day throughout the week. \\
	\textbf{1. Absolute POI Act Observation:} 
	From a single temporal aspect, Absolute POI Act Observation clarifies that many of POIs are frequently visited either during $w_d$ or $w_e$. Therefore, for each $p_j$, visited by a set of users $U_j$, we computed $p_j^{a*}$ (Eq. \ref{POI_Act_mrgin_Equation}) as an absolute rate of temporal $w_d$/$w_e$ deviation. In the inspection, we chose those locations which were visited by at least 5 users.		
	\begin{equation}
	\small
	\label{POI_Act_mrgin_Equation}
	p_j^{a*} = \frac {\sum_{u_i \in U_j}|p_{i,j}^d-p_{i,j}^e|}{|U_j|} 
	\end{equation}
	$p_{i,j}^d$ and $p_{i,j}^e$ are the probabilities of each $u_i \in U_j$ to visit $p_j$ during $w_d$ and $w_e$(Eq.\ref{eq:UPOI_ACT}):
	\begin{equation}
	\small
	\label{eq:UPOI_ACT}
	p_{i,j}^d=\frac{W_{i,j}^d}{W_{i,j}}, p_{i,j}^e=\frac{W_{i,j}^e}{W_{i,j}} 
	\end{equation}
	    Here $W_{i,j}$ is the total number of times that each $u_i \in U_j$ has visited $p_j$. Also, $W_{i,j}^d$ and $W_{i,j}^e$ record the number of visits performed during $w_d$ and, respectively.\\
	    The probabilities regarding locations' weekly deviations in various ranges (e.g. 0.3 - 0.4) are displayed in figures \ref{POI_Act_Observation}(a) and (b). This shows that more than 70\% of the venues in both datasets gain an average absolute orientation greater than threshold $T$. As highlighted in dark orange, we observe that most of the locations in both datasets are mainly visited either during the weekend or weekday.\\
	\textbf{2. Absolute User Act Observation:} Similar to POI Act Observation, considering single weekday/weekend cycle, we computed $u_i^{a*}$ (Eq. \ref{User_Act_mrgin_Equation}) as the average rate of absolute temporal $w_d$/$w_e$ deviation for each user $u_i$ with $L_i$ as her visiting record. We selected users who have visited at least 8 POIs ($\{\forall u_i \in \mathbb{U} | \left| L_i \right|>8\}$). Figures \ref{UserAct_Observation}(a) and \ref{UserAct_Observation}(b) illustrate relevant probabilistic bins which reflect to what extent each user is temporally oriented toward either $w_d$ or $w_e$. $\left|p_{i,j}^a\right|$ is $p_j$'s absolute POI act limited to $u_i$'s visits.
	\begin{equation}
	\small
	\label{User_Act_mrgin_Equation}
	u_i^{a*} = \frac {\sum_{p_j \in L_i}|p_{i,j}^a|}{|L_i|} 
	\end{equation}	
    If $u_i^{a*}$ is less than $T$(15\%), we can ensure that  $u_i$ is not oriented toward $w_d$ or $w_e$. However, as highlighted in dark orange (Fig. \ref{UserAct_Observation}), 57.3\% and 61.6\% of users in Foursquare and Brightkite have an absolute temporal deviation more than the $T$. Also, more than 10\% of users are highly aligned toward either weekday or weekend ($u_i^{a*}>45\%$).\\
    Based on aforementioned observations, we can conclude that weekly temporal influences exist for both users and POIs. However, we have only taken a single temporal aspect (i.e. the weekly cycle) into consideration.
	\subsection{User Act Single Factor Model}
	\label{User_Act_Efficient_Model}
	As shown in observations, the effect of single temporal influence regarding weekly periods has been witnessed in the visibility patterns of users and POIs. Hence, in this section, we review the method which was used in our previous work (\cite{Hosseini2016}) to compute user acts. In reality, primary user act (Def. \ref{def:USER_ACT}, Eq. \ref{eq:USER_ACT}) treats all the locations the same, while they differ based on their importance (\textit{Weekday/Weekend visiting influence}). Therefore, we propose a more efficient model to compute the user act. We first need to obtain user's visiting orientation toward $w_d$ or $w_e$.  Therefore, we compute the POI act for every location visited by $u_i$ ($p_j \in L_i$). We use Eq. \ref{eq:pijlambda} to find positive or negative impacts.
	\begin{equation}
	\small
	\label{eq:pijlambda}
	\hat{p}_{i,j}^d=(p_{i,j}^d - \lambda) , \hat{p}_{i,j}^e=(p_{i,j}^e - \lambda)
	\end{equation} 
	Where $\lambda \in (0,1)$ distinguishes $w_d$/$w_e$ margins. If we assume $\lambda=0.5$, $p_{i,j}^d= 0.75$ and $p_{i,j}^e=0.25$, then $\hat{p}_{i,j}^d=0.75-0.5=0.25$, which indicates that $p_j$ has a positive impact on user $i's$ weekday act. For computation of a user's act, we assume that as higher the visiting probability is the location will be more significant in the final value of the user's act. In our previous work we utilized \cite{Ye2011} to compute non-temporal visiting influences for each location ($p_j$). we removed each $p_j$ from $L_i$, we then obtained $c^*_{i,j}$ which represents the probability of $u_i$ to visit $p_j$ comprising all modules of Collaboration, Friendship, and Vicinity. We then normalized the results ($\in$ (0,1]) using feature scaling (Eq. \ref{eq:CIJ}):
	\begin{equation}
	\label{eq:CIJ}
	\hat{c}^*_{i,j}=\frac{c^*_{i,j}-Min_{ci}}{Max_{ci}-Min_{ci}},
	\end{equation}
	where $Max_{ci}=arg_{max} ({c^*_{i,k}}), Min_{ci}=arg_{min} ({c^*_{i,k}}), \forall p_k \in L_i$. 
	To get the final weekday orientation probability for each $p_j \in L_i$ we  use Eq. \ref{eq:DAYIJ}:
	\begin{equation}
	\small
	\label{eq:DAYIJ}
	Pr_{i,j}^d=\hat{c}^*_{i,j} * \hat{p}_{i,j}^d=\frac{c^*_{i,j}-Min_{ci}}{Max_{ci}-Min_{ci}} * (p_{i,j}^d - \lambda) 
	\end{equation}
	The higher $\hat{c}^*_{i,j}$ is, the more likely this location will be visited by $u_i$ and will be more influential on $u_i$'s act. Similarly, the weekend orientation probability ($Pr_{i,j}^e$) can be computed as follows:
	\begin{equation}
	\small
	\label{eq:ENDIJ}
	Pr_{i,j}^e=\hat{c}^*_{i,j} * \hat{p}_{i,j}^e=\frac{c^*_{i,j}-Min_{ci}}{Max_{ci}-Min_{ci}} * (p_{i,j}^e - \lambda) 
	\end{equation}
	Finally, the user act orientation based on the single temporal influence is obtained through Eq. \ref{eq:AVGFinal}:
	\begin{equation}
	\small
	\label{eq:AVGFinal}
	\hat{u_i^a}=\left |\tilde{Avg}_i^d-\tilde{Avg}_i^e \right |
	\end{equation}
	While $\tilde{Avg}_i^d$ (Eq. \ref{eq:AVGD}) and $\tilde{Avg}_i^e$ (\ref{eq:AVGE}) are respective $w_d$/$w_e$ average ratios.
	\begin{equation}
	\small
	\label{eq:AVGD}
	\tilde{Avg}_i^d=\frac{\varSigma_{p_j\in L_i}Pr_{i,j}^d}{|L_i|}
	\end{equation}
	\begin{equation}
	\small
	\label{eq:AVGE}
	\tilde{Avg}_i^e=\frac{\varSigma_{p_j\in L_i}Pr_{i,j}^e}{|L_i|}
	\end{equation}
	If the value of $\tilde{Avg}_i^d$-$\tilde{Avg}_i^e$ is greater than zero, this will denote that the user is aligned toward $w_d$ and if it is less than zero, it means $w_e$ orientation.		
	\subsection{Uni-variate Temporal Framework}
Univariate Temporal Framework developed in our recent work (\cite{Hosseini2016}) proposes a ranked list of candidate POIs for the query user solely depending on the extent of her univariate (weekday/weekend) temporal orientation.
	\begin{figure}[!htp]
		\tiny
		\centering
		\includegraphics[width=0.45\textwidth]{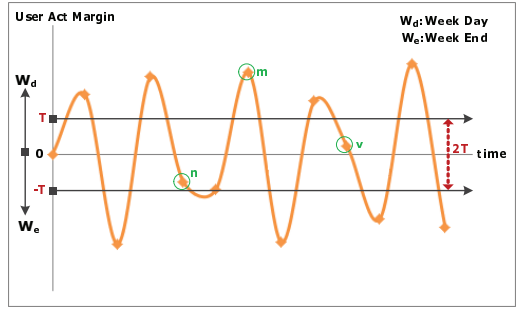}
		\caption{\small Continuous stream of users is discretized by first computing relevant user acts and then utilizing of threshold $T$.}
		\label{fig:ThresholdDiagram_fig}
	\end{figure}
    As Fig. \ref{fig:ThresholdDiagram_fig} depicts, we can imagine the input of a recommender system as a continuous stream of query users in the course of time. Utilizing check-in history, the system aims to find top \textit{@Num} unvisited venues for each user. A basic POI recommender system doesn't differentiate $w_d/w_e$ temporal preferences, however, we used threshold $T$ as a single temporal influence margin to discretise stream of query users based on their effective user acts (Section \ref{User_Act_Efficient_Model}). If they exceed the threshold, univariate temporal influence will be employed to enhance the results. Otherwise, they will be proposed with non-temporal influences. For example, $u_m$ and $u_v$ are oriented to do the check-ins during $w_d$. However, unlike $u_m$, $u_v$ doesn't surpass $T$ and is not adequately oriented toward $w_d$ so the framework doesn't apply the temporal method for her. While the user act reflects how a user performs the check-ins in weekly cycles, POI act is used in recommendation process to suggest right POIs to the right users through utilizing of single slot time-related effect.
	\begin{equation}
	\small
	\label{eq:Framework}
	f(L_i)= 
	\begin{cases}
	M_{avg}(\rho,\delta)& \text{if } \hat{u_i^a}\geq T\\
usg_w     & \text{otherwise}
\end{cases}
\end{equation}
As formulated in Eq. \ref{eq:Framework}, the system receives $u_i$'s check-in history ($L_i$). If the user act computed based on $L_i$ exceeds threshold $T$, the system will utilize temporal influence. Otherwise, the user will be recommended by $usg_w$ (\cite{Ye2011}). In temporal case, $\rho$ is the initial list of recommending POIs computed by USG and $\delta$ resembles the POI act for each item in the primary recommendation list. $\rho$ and $\delta$ are the input of $M_{avg}$ function which performs recommendation as described in section \ref{Temporal_Act_Based_Recommendation}.
\subsection{Uni-variate Temporal Recommendation}
\label{Temporal_Act_Based_Recommendation}
If the user act is greater than threshold ($\hat{u_i^a} \geq T$), we need to follow univariate temporal recommendation approach ($M_{avg}$).The method has two inputs. $\rho$ which is the primary decently sorted recommendation list and $\delta$ which includes the univariate temporal acts for each of POIs in $\rho$. We first retrieve top \textit{K*@Num} items from $\rho$ where \textit{@Num} is the number of final list (denoted as $R$). $R$ is formed by three subsets of Weekday aligned($R^d$), Weekend oriented($R^e$) and Neutral ($R^n$) where $R=\{R^d,R^e,R^n\}$ and $|R|=@Num$. The final proportion of each category will follow relevant ratios from proper POIs which are computed based on efficient user act(Eq. \ref{eq:RecommentationRanks}).
\begin{equation}
\small
\label{eq:RecommentationRanks}
M_{avg}(\rho,\delta)= 
\begin{cases}
\left|R^d\right|=(\tilde{Avg}_i^d+\lambda-\frac{\xi}{2})* @Num & \text{if } p_y^a>\theta\\
\left|R^e\right|=(\tilde{Avg}_i^e+\lambda-\frac{\xi}{2})* @Num & \text{if } p_y^a<\theta\\
\left|R^n\right|=\xi* @Num & Otherwise
\end{cases}
\end{equation}
$(\tilde{Avg}_i^d+\lambda-\frac{\xi}{2})$ and $(\tilde{Avg}_i^e+\lambda-\frac{\xi}{2})$ are respective $w_d$ and $w_e$ proportions from final recommendation list. Also, $\theta$ is the threshold to distinguish $w_d$/$w_e$ oriented POIs. For example, if $\theta=0$, the weekday portion from the final list will comprise the POIs whose acts are greater than 0 ($\forall p_y^a \in \delta|p_y^a>0$) and for weekend ratio the POI acts should be less than 0 ($\forall p_y^a \in \delta|p_y^a<0$). In fact, Neutral POIs are not likely to have high scores in $w_d$/$w_e$ lists. However, we still need to propose them when they gain high probabilities. Therefore we reserve a minor portion ($\xi$) for POIs which are not temporally aligned.
\section{Multi-aspect Time-related Influence}
\begin{figure}
	\tiny
	\centering
	\includegraphics[width=0.45\textwidth]{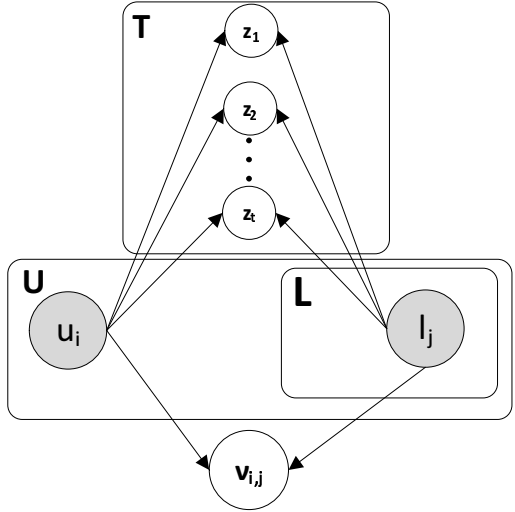}
	\caption{\small The graphical representation of Multi-aspect Time-related Influence in location recommendation.}
	\label{fig:Latent_Diagram}
\end{figure}

Joint with collaborative filtering methods, effects such as \textit{Geographical} (\cite{Ye2011},\cite{Yuan2013},\cite{Liu2013}), \textit{social} (\cite{Cheng2012},\cite{Gao2012}) and  \textit{context-oriented} including text contents and word-of-mouth (\cite{Yin2013,Yin2015}) are already employed to improve the effectiveness of spatial item recommendation. A growing line of research has also utilized \textit{temporal influence} to foster the same purpose. However, the majority of prior works merely consider univariate temporal granularities like hour of the day (\cite{Yin2016,Gao2013,Yuan2013,Fang2016,Zhang2015b,Deveaud2015}), day of the week or weekday/weekend cycles (\cite{Zhao2016,Gao2013,Yuan2013a,Zhang}). Anyhow, selecting one temporal dimension and leave others unattended is problematic, even if it is owing to complexity or overfitting issues (\cite{Zhang2015b}).\\ Practically, LBSN based location recommendation systems consider bigger granularities such as the hour, day, week owing to sparsity issues. However, in other dense datasets suchlike user-item feedback matrices generated in online social networking spheres (e.g. Facebook), smaller granularities can be taken into account to study users' mobility behaviors more precisely. In fact, all temporal granularities follow the \textit{Temporal Subset Property} (TSP) which means some of the time slots are the subset of the others($minute \subset hour \subset day \subset week $). 
Hence, rather than reconfiguring a method (\cite{Gao2013}) to make it work using another single time slot, it is better to develop an approach which can include multiple temporal factors in a unified way.
In short, we believe there is a multivariate compound temporal correlation between visibility patterns of LBSN users and locations. Inspired by this perception, we propose our Bayesian model (Fig. \ref{fig:Latent_Diagram}) which is capable of employing an infinite number of temporal scales in location recommendation systems which we name it \textit{MATI}, standing for \textit{Multi-aspect Time-related Influence}.
As illustrated in figure \ref{fig:Latent_Diagram}, each user $u_i \in U$ can visit any location $l_j \in L$ affected under certain constraints that we can categorize them into \textit{multiple temporal latent factors} defined as $T=\{z_1,z_2, \dots ,z_t\}$ and \textit{non-temporal impacts} (denoted as $\nu_{i,j}$). $\nu_{i,j}$ represents the visibility impact of the user $u_i$ to visit location $l_j$ disregarding the temporal influence. Such an impact is involved with friendship, geographical, and context-oriented influences.\\
In order to clarify our method, we demonstrate latent temporal parameters using two scales of $z_h$ regarding time of the day and $z_d$ for the day of the week. Yet, our model can be generalized to dealing with multiple time-related granularities as proven in section \ref{Parameter_Inference_Algorithm}.\\
Apparently, LBSN users mostly own a limited number of check-ins which results in a sparse user-POI matrix. Adding the time dimension provokes User-Time-POI (UTP) cube that is even more dispersed. While User-POI matrix reports whether a user has visited a location or not, UTP cube further provides the time information. In reality, UTP cubes can be defined in various levels of granularity where each scale is associated with a temporal latent factor.\\
As our approach aims to utilize multiple temporal factors to improve the effectiveness of location recommendation systems, we propose our method for sparseness demotion and exploiting of temporal slabs in section \ref{Exploiting_Temporal_Blocks} and continue with the parameter inference algorithm in section \ref{Parameter_Inference_Algorithm}.
\subsection{Exploiting multi-aspect temporal slabs}
\label{Exploiting_Temporal_Blocks}
Each temporal scale owns a distinguished User-Time-POI matrix as well as an associated level of sparsity.
The UTP cube is certainly dispersed and the level of sparsity has an indirect connection to the size of discretised temporal granularity. For instance, associated UTP cube with the minute interval is far more scattered than hourly time slot. We assign a latent factor for each temporal granularity.
\begin{definition}
	\label{def:Latent_Temporal_Factors}
	\textbf{(Latent Temporal Factors)} Given a set of users $\mathbb{U}=\{u_1, u_2,\dots,u_n\}$ and POIs $\mathbb{P}=\{p_1, p_2,\dots, p_n\}$, we define that each $u_i$ ($\forall u_i \in \mathbb{U}$) can visit a location $p_j$ under a predefined set of Latent Temporal Factors $\mathbb{T}=\{z_1,z_2, \dots ,z_t\}$.
\end{definition}
Referring to definition \ref{def:Latent_Temporal_Factors}, if there is only one latent factor merely defined for hour (i.e. $\mathbb{T}=\{z_h\}$), then $u_i$ will own 24 vectors. Each of them will report the locations that he has visited during every 24 hours of day/night cycle. As a matter of fact, check-in activity of the user during certain hours will be similar (\cite{Yuan2013}). Using the vectors, we can apply Cosine or Pearson (with similar results) metric to compute the similarity between each pair of hours. The final similarity value among a twin will be the average value that is gained from each of the users who have performed the check-ins during both hours. Intuitively, if two hours are similar, a check-in during each of them can also be counted for the other one. Therefore, we can combine similar hours and make a block of hours that we name it the \textit{uni-aspect temporal slab}.
\begin{definition}
	\label{def:uniaspect_temporal_slab}
	\textbf{(Uni-Aspect Temporal Slab)} Given a latent temporal factor $z_h$ comprising $m$ default intervals (e.g. 24 hourly slots) $z_h=\{c_1^h,c_2^h, \dots c_m^h\}$, we can construct a sample \textit{Uni-Aspect Temporal Slab} $z_h^i$ through merging similar slots from $z_h$'s intervals. 
\end{definition}
For instance, considering $z_h$ as the hour latent factor, $z_h^i=\{21,22,23\}$ can be the hourly slab which is made up of 3 hours (i.e. from 9pm,10pm, and 11pm). Nevertheless, this is a one dimension temporal slab. If we consider two latent features regarding hour and day (i.e. $\mathbb{T}=\{z_h,z_d\}$). Initially, we can witness the subset feature (TSP: $z_h \subset z_d$). While we have already exploited distinguished uni-aspect slabs w.r.t. $z_h$ and $z_d$, we can imagine a multi-aspect temporal slab $\tau_o^s$ as a combination of two vectors of $z_h^j$ and $z_d^k$ regarding respective hourly and daily slabs.\\
\begin{definition}
	\label{def:multiaspect_temporal_slab}
	\textbf{(Multi-Aspect Temporal Slab)} Given the set of Uni-Aspect Temporal Slabs extracted for $n$ latent factors, a Multi-Aspect Temporal Slab $\tau_o^s$ is formed via combining $n$ Uni-Aspect Temporal Slabs where each of them is retrieved from a distinguished latent factor.
\end{definition}
We can now formulate the problem regarding extraction of multi-aspect temporal slabs
\begin{problem}
	\label{def:Exploiting_multi_aspect_temporal_slabs}
	\textbf{(Exploiting Multi-Aspect Temporal Slabs)} 
\textit{Given a set of predefined temporal latent factors ($\mathbb{T}$) as well as the set of users ($\mathbb{U}$) and their check-in logs $\mathbb{L}=\{L_1, L_2,\dots,L_n\}$ (e.g. $L_1$ is the visiting history of user $u_1$), our goal is to extract all possible Multi-Aspect Temporal Slabs.}
\end{problem}
We aim to reduce sparsity in UTP cubes through extracting multi-aspect temporal slabs (Def. \ref{def:multiaspect_temporal_slab}). We first need to leverage uni-aspect temporal slabs (Def. \ref{def:uniaspect_temporal_slab}) that are constructed through computing the similarity between each pair of temporal slots for every latent time-related feature. Due to sparsity, finding the similarity value between two temporal slots is also challenging. Hence, in this section, we propose our method to solve the Prob.\ref{def:Exploiting_multi_aspect_temporal_slabs}. For clarity purposes, we provide the solution for two factors of $z_h$ and $z_d$ (Hours of the day, day of the week) however, the same method can be applied for multiple latent features.\\
Considering latent factors with smaller scales (e.g. minutes) the sparsity condition may get even worse. Random sampling of the check-in log owned by a portion of dataset users and relying on the average similarity values to compute the final metric between two temporal slots is the first solution. However, it suffers from two pitfalls. Firstly it cannot provide a comprehensive picture of the entire data set. Secondly, the sample number of users may miss providing the similarity among some of the slots. Owing to the common limited number of check-ins performed by LBSN users, given a sample user $u_i$, she might not have generated a check-in at all hour slots of a day/night ($z_h$) nor every day in week cycles($z_d$). Therefore, we consider an alternative approach which utilizes an iterative process that takes random $n\%$ of the users in each round and subsequently computes similarity values for each chosen user. We employ non-replacement stratified sampling model (\cite{Liberty2016}). This method splits the data into several partitions based on the variety of users including passive, semi-active, and active and subsequently draws random samples from each partition. We then repeated the procedure until we collected a minimum of $m$ similarity samples between each pair of slots for selected latent factors. While the similarity pairs are more reliable, this finally obtains a better view of the whole dataset \footnote{We created matrices of h*h using LINQ queries in which h is the number of slots in each temporal scale (i.e. 7 for $z_d$ and 24 for $z_h$)}. Exceptionally, for the small datasets (i.e. Foursquare), even after completing the sampling process on all of the training pilot set (80\% for four iterations), the entries in both of similarity matrices (regarding $z_d$ and $z_h$) were either incomplete or unreliable. In order to predict missing similarity values between some of the temporal slots, we utilized matrix factorization(\cite{Goyal2010}). Nonetheless, Normal Equation through a regression problem could be another choice.\\
\begin{figure}[!htb]
	\centering
	\minipage{0.25\textwidth}
	\centering
	\includegraphics[width=\linewidth]{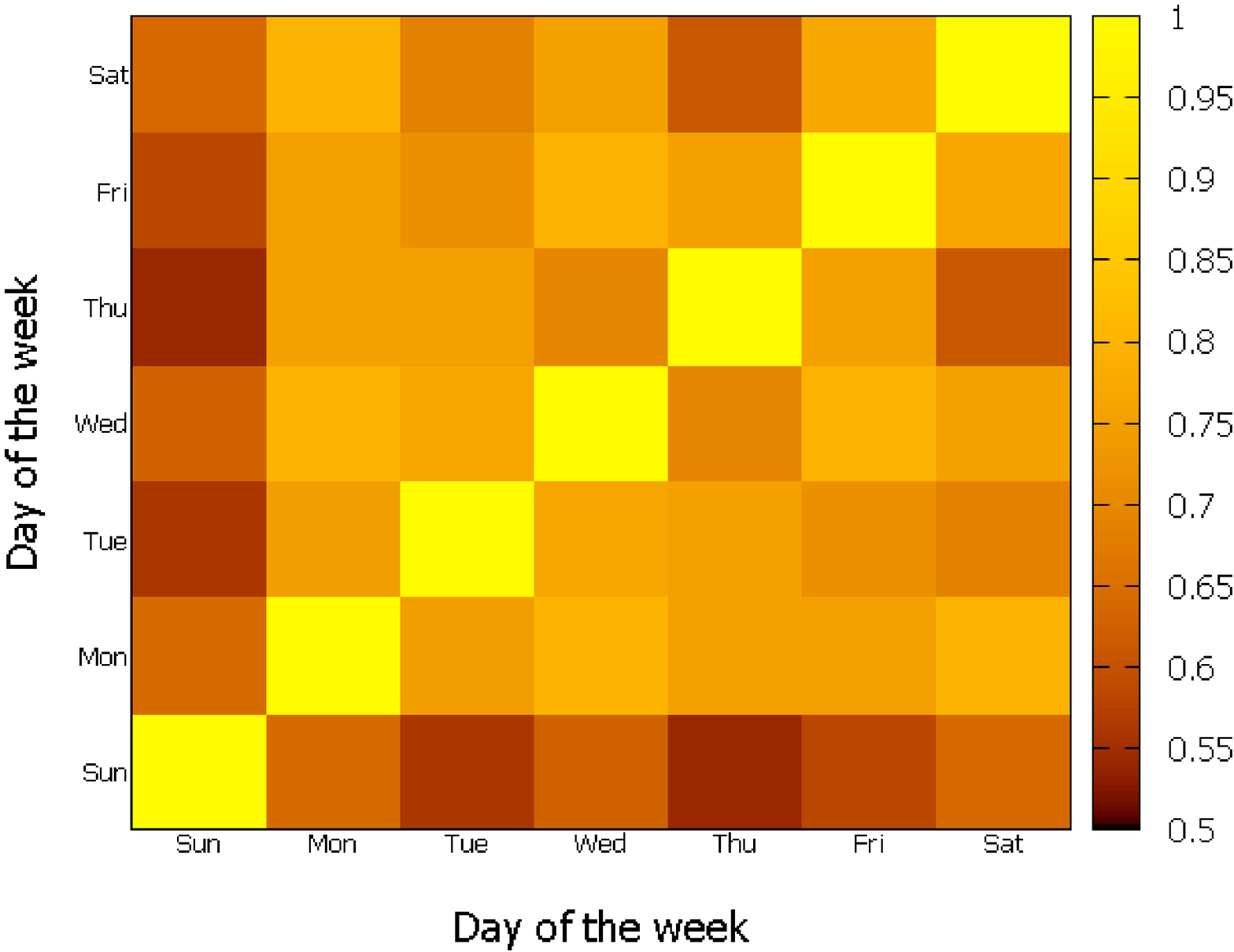} 
	\small (a) Foursquare: Day ($z_d$)
	\endminipage\hfill
	\minipage{0.25\textwidth}
	\centering
	\includegraphics[width=\linewidth]{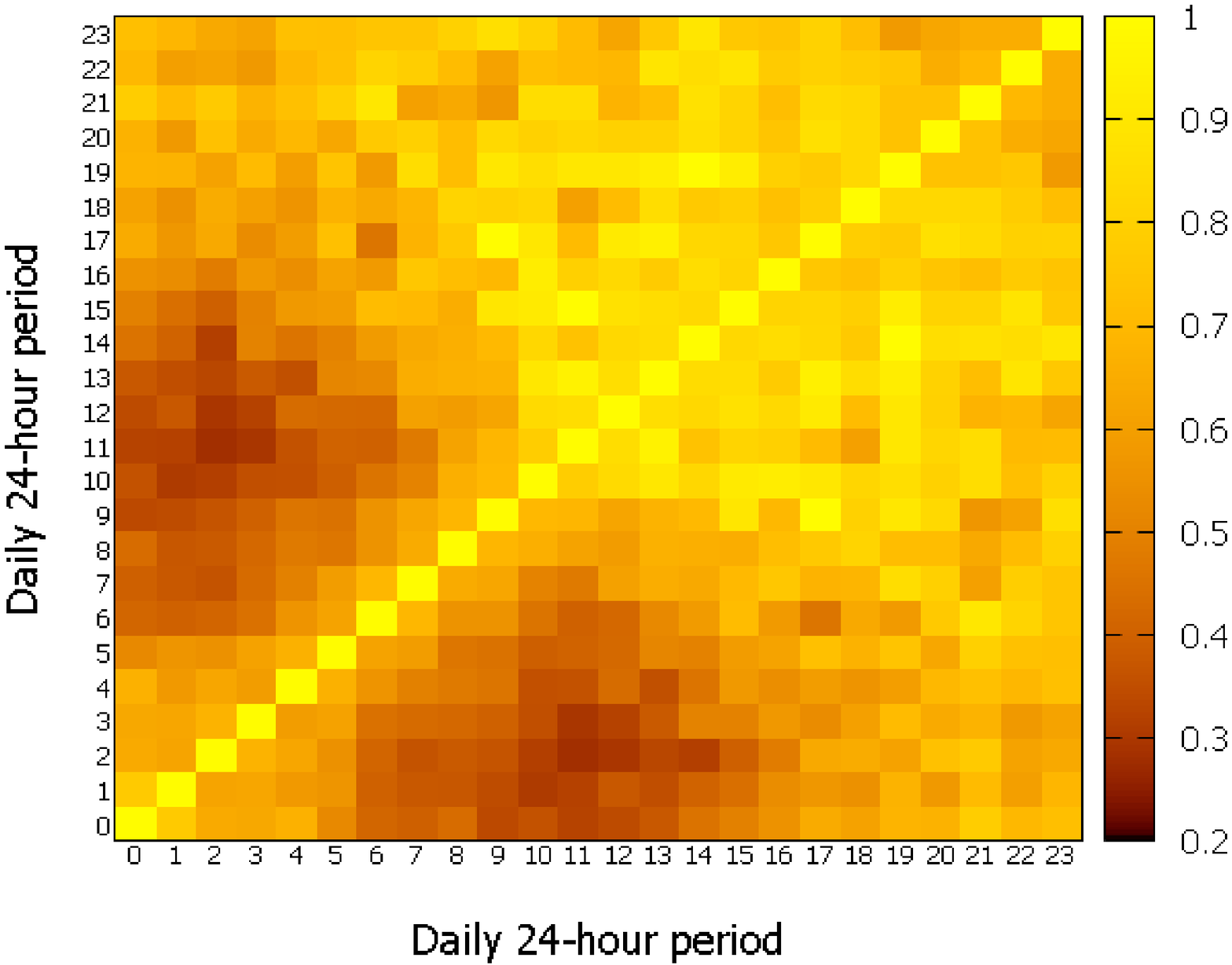} 
	\small (b) Foursquare: Hour ($z_h$)
	\endminipage\hfill
	\minipage{0.25\textwidth}
	\includegraphics[width=\linewidth]{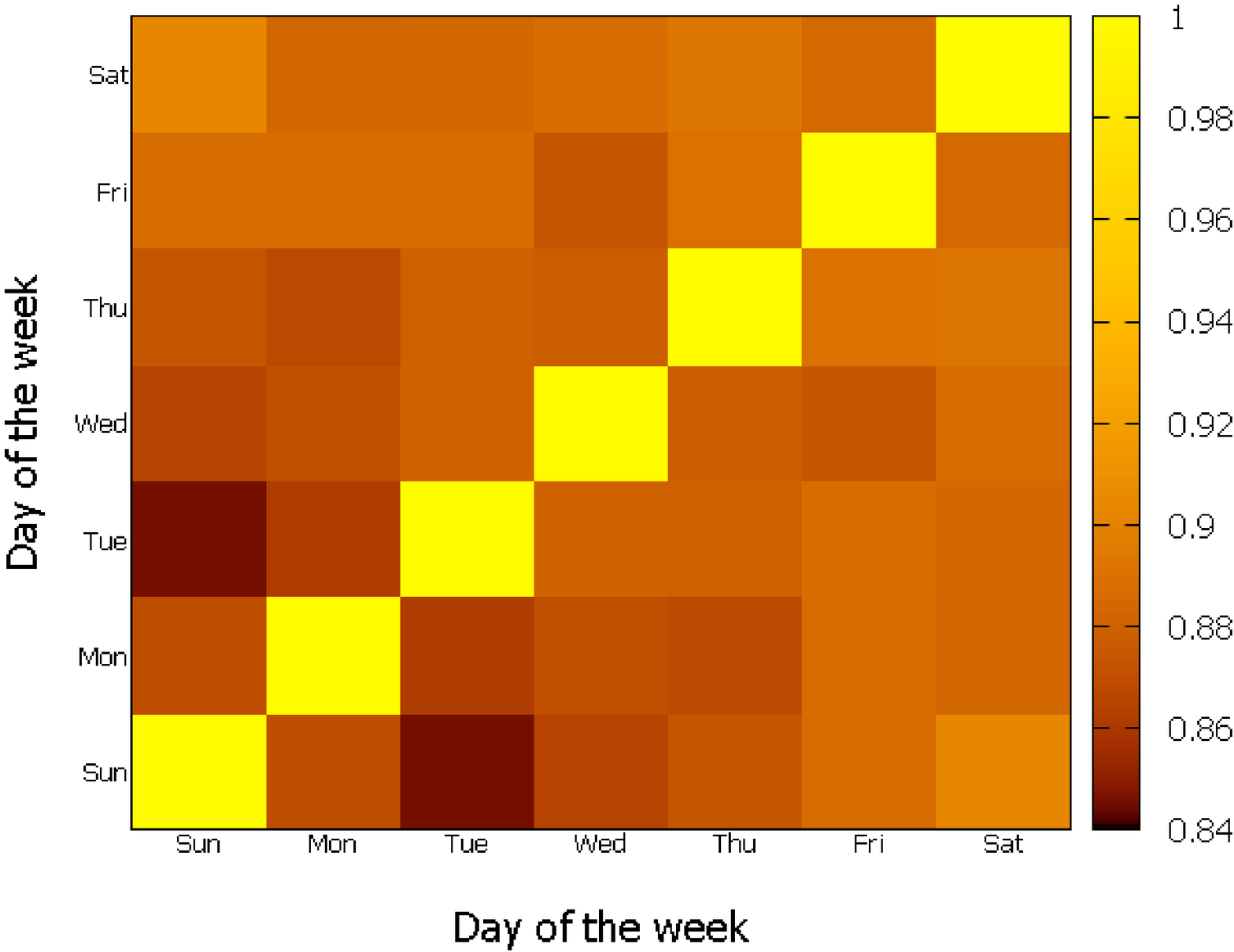} 
	\small (c) Brightkite: Day ($z_d$)
	\endminipage\hfill
	\minipage{0.25\textwidth}
	\centering
	\includegraphics[width=\linewidth]{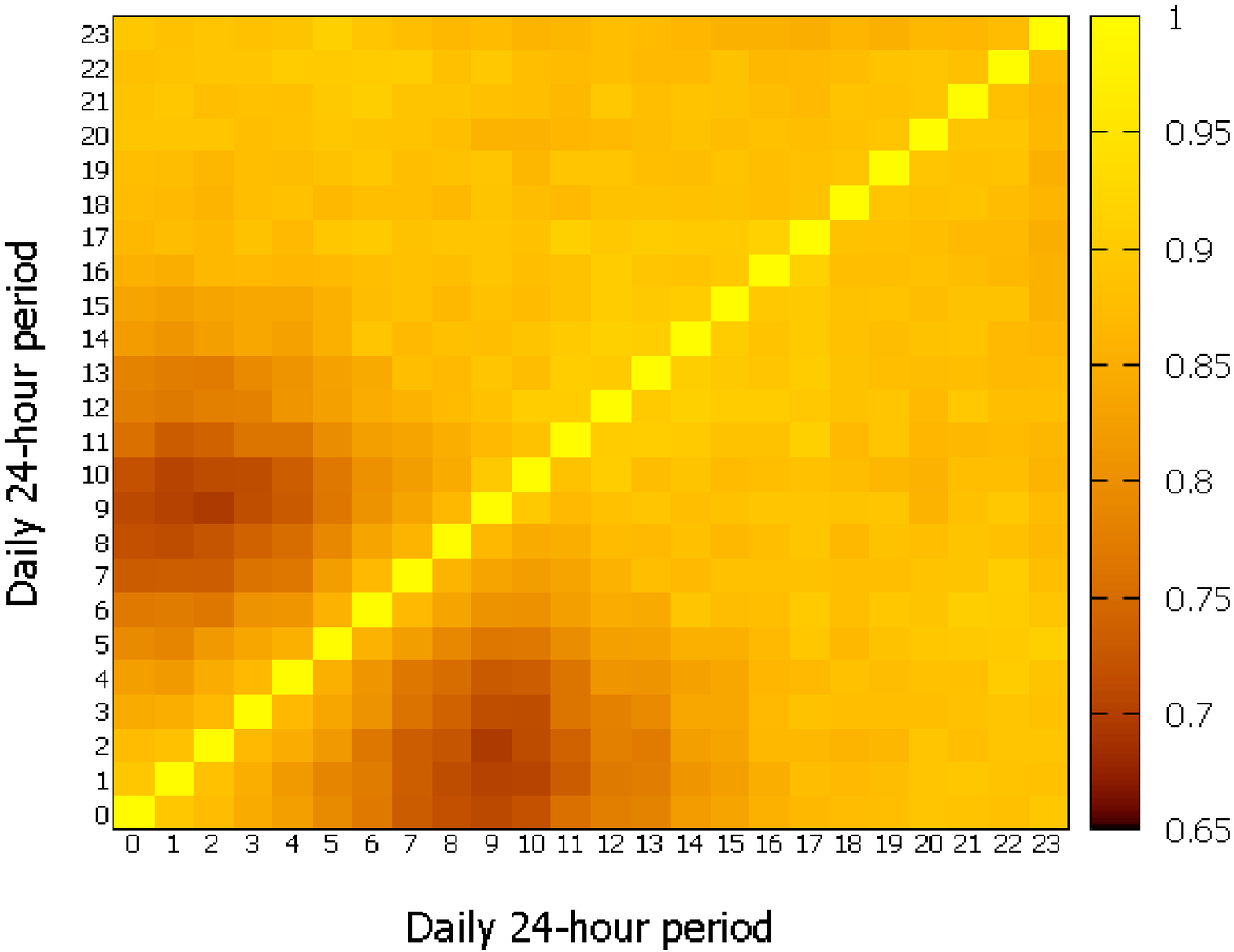} 
	\small (d) Brightkite: Hour ($z_h$)
	\endminipage
	\caption{Similarity between temporal slots}
	\label{fig:Similarity_Slots}
\end{figure}
Figure \ref{fig:Similarity_Slots} illustrates the similarity maps associated with twin latent variables in our both datasets.
The primary discoveries in figure \ref{fig:Similarity_Slots} are three-fold: (i) Neighboring temporal slots are more similar which affirms observations performed by Yuan et al. (\cite{Yuan2013}). (ii) Although, due to the high sampling rate of $z_h$ factor, Brightkite's similarity map is more smooth. However, the hour based ($z_h$) similarity patterns in both datasets (Figures \ref{fig:Similarity_Slots}(b), \ref{fig:Similarity_Slots}(d)) are quite similar. (iii) While the similarity map regarding $z_h$ in Brightkite dataset is similar to the Foursquare counterpart, the figures regarding $z_d$ (\ref{fig:Similarity_Slots}(a), \ref{fig:Similarity_Slots}(c)) are obviously different. This implies that on the one hand, including more latent parameters can better reveal time-oriented mobility patterns and on the other hand a single temporal scale can not be applicable to all datasets.\\
Nevertheless, in order to generate uni-aspect temporal slabs regarding each latent feature, we were required to partition adjacent similar slots through various distance thresholds. We had to consider scalability features in mind to eventually mitigate the sparsity impact. Hence, we opted for bottom-up Hierarchical Agglomerative Clustering (\textit{HAC}) which have gained a good reputation in maximizing similarity \cite{Das2014}. HAC employs a likeness function to assure that similar pairs will be included in the same cluster. We used \textit{complete linkage} to ensure that all the time slots inside each of merging clusters have similar visiting patterns.\\
\begin{figure}[!htb]	
	\centering
	\minipage{0.13\textwidth}
	\includegraphics[width=\linewidth]{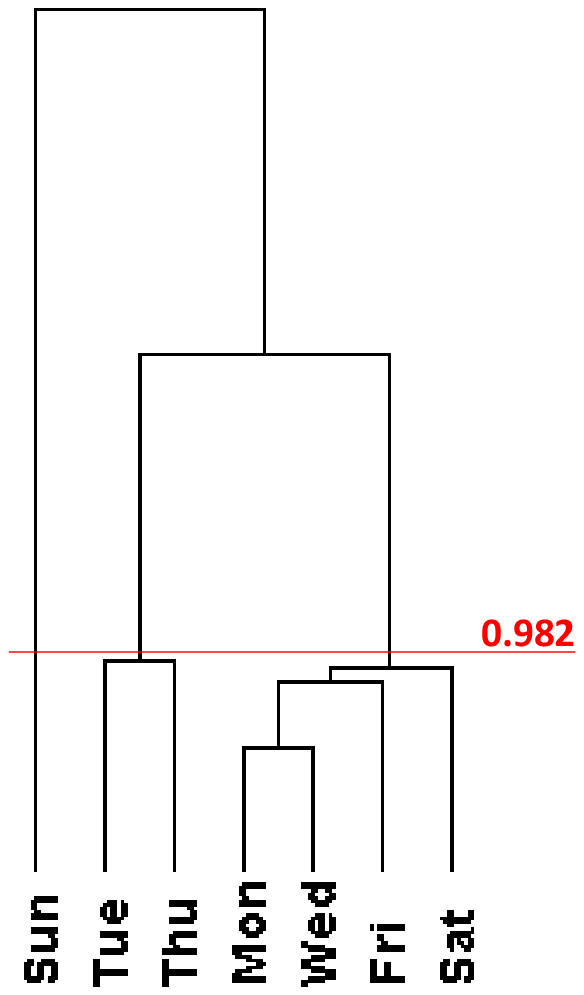} 
	\small (a) Foursquare:\\$z_d$
	\endminipage\hfill
	\minipage{0.37\textwidth}%
	\centering
	\includegraphics[width=\linewidth]{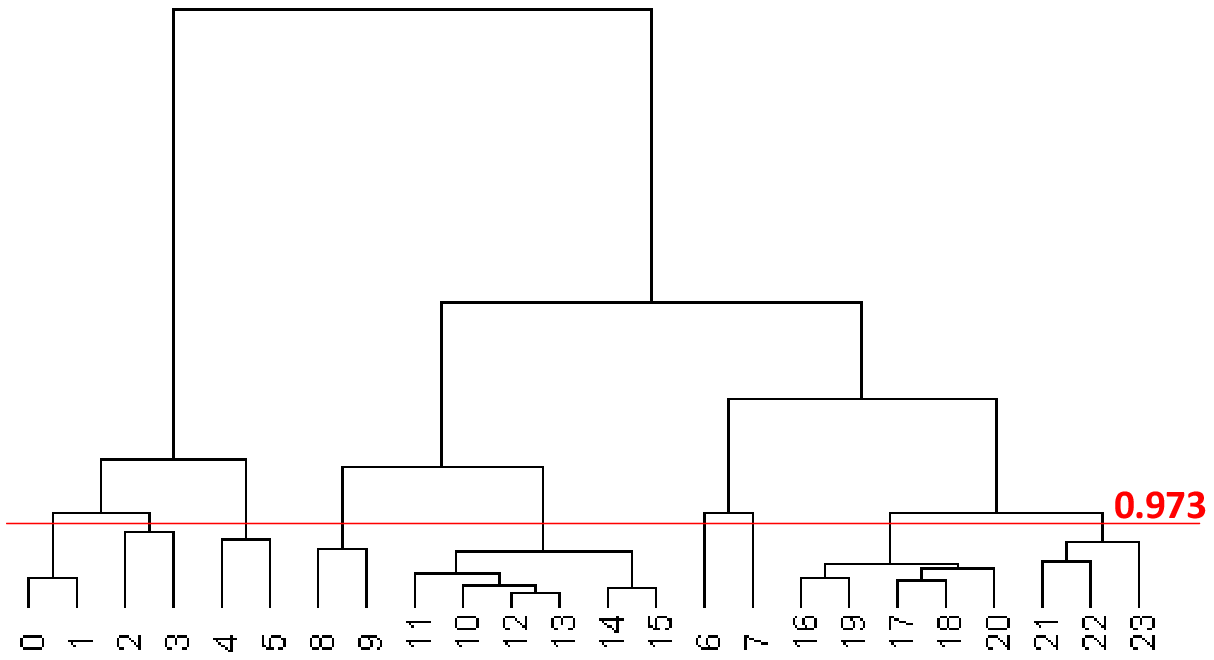} 
	\small (b) Foursquare: $z_h$
	\endminipage\hfill
	\minipage{0.13\textwidth}
	\centering
	\includegraphics[width=\linewidth]{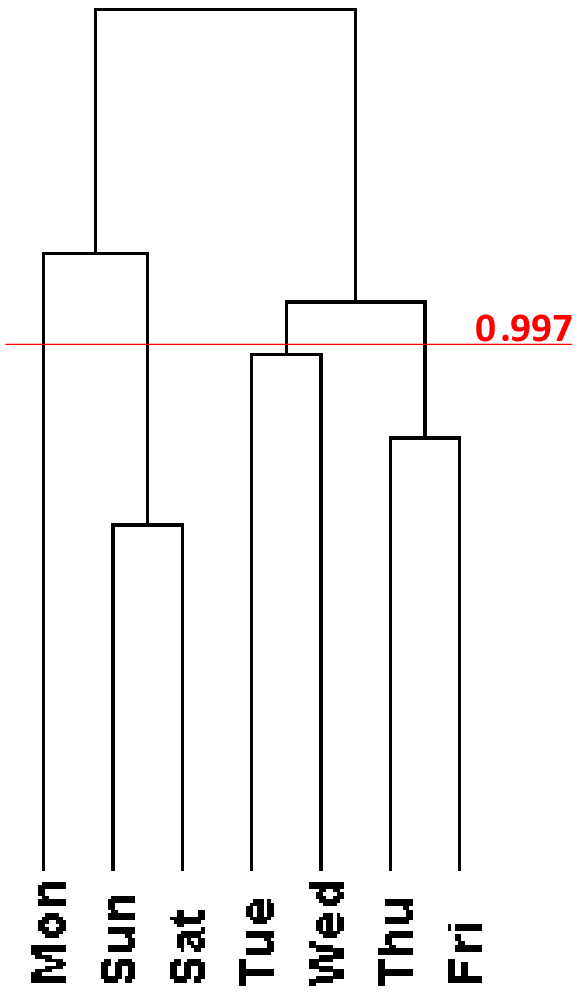} 
	\small (c) Brightkite: $z_d$
	\endminipage\hfill
	\minipage{0.37\textwidth}%
	\centering
	\includegraphics[width=\linewidth]{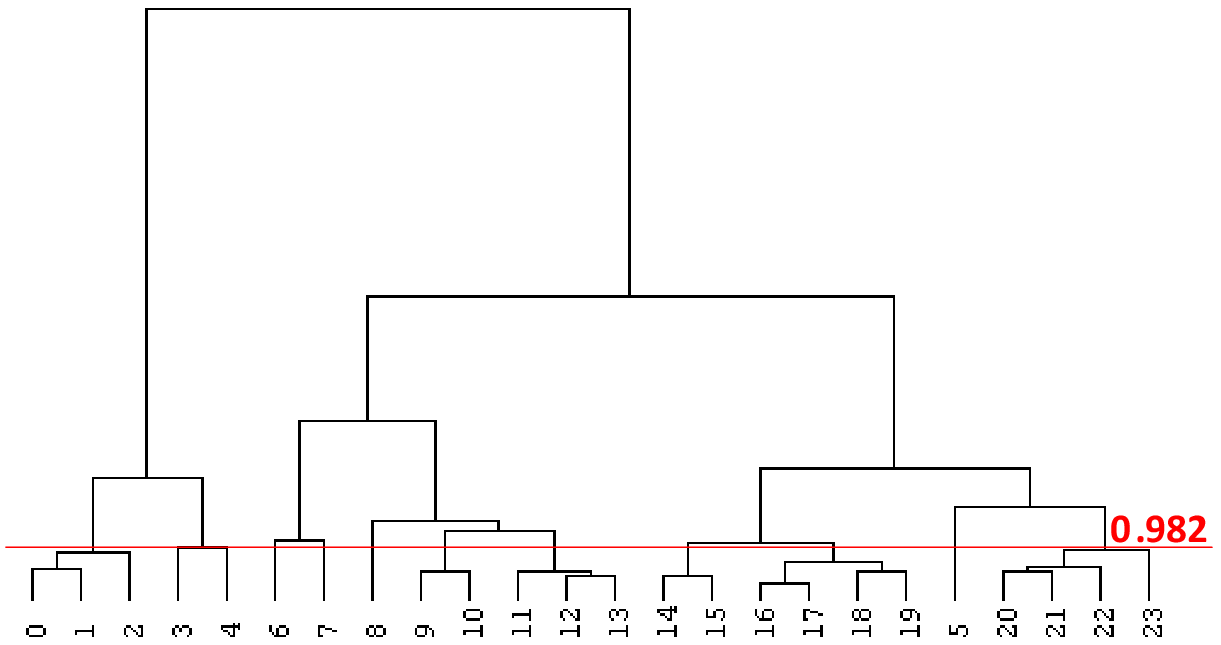} 
	\small (d) Brightkite: $z_d$
	\endminipage	
	\caption{Exploiting Temporal Slabs}
	\label{fig:HAC_Slabs}
\end{figure}
Figure \ref{fig:HAC_Slabs} exhibits the dendrogram regarding each of the latent variables in both datasets.The red line demonstrates the threshold value corresponding to the correlation among complete linkage. A final multi-aspect temporal slab will have the merging vertices regarding both temporal parameters. For example, as figure \ref{fig:HAC_Slabs} (a) shows, Tuesday and Thursday are similar enough to be merged as a $z_d$ block in the foursquare dataset. In addition, figure \ref{fig:HAC_Slabs} (b) indicates that three hours of 21, 22, and 23 can be merged into a $z_h$ uni-aspect temporal slab. Hence, we can denote $\tau_i^s$ as an independent \textit{multi-aspect temporal slab} that has two vector attributes, representing both latent aspects of $z_d^r=\{Tuesday,Thursday\}$ and $z_h^v=\{21,22,23\}$.
Now that we have proposed the solution to exploit multi-aspect temporal slabs, we can formulate the main problem as follows:
\begin{problem}
	\label{Recommendation_multi_aspect_temporal_Slabs}
	\textbf{(Recommendation Via Multi-Aspect Temporal Slabs)} 
	\textit{Given the check-in log dataset $\mathbb{D}$, a predefined set of latent temporal factors $\mathbb{T}$, set of exploited multi-aspect temporal slabs $\mathbb{\tau}^s$ based on $\mathbb{T}$ and the query user $u_i$, our goal is to suggest a list of new POIs that $u_i$ would likely visit, while proposed locations are correlated with $u_i$ according to $\left| \mathbb{T} \right|$ temporal aspects.}
\end{problem}
\subsection{Parameter Inference Algorithm in Recommendation}
\label{Parameter_Inference_Algorithm}
The problem of suggesting unvisited places to a user $u_i$ can be undertaken by computing the probability for $u_i$ to visit a spatial item $l_j$ denoted as $Pr(l_j|u_i)$ and formulated by Eq. \ref{Step_1}. While, all users are of the same importance, a set of highly ranked locations will be proposed to $u_i$.
\begin{equation}
\small
\label{Step_1}
Pr(l_j|u_i)=\frac{Pr(u_i,l_j)}{Pr(u_i)} \propto Pr(u_i,l_j)
\end{equation}
According to the MATI model, we can include as many latent temporal variables as required. Alongside other impacts (i.e. Geographical, Social and Context-oriented influences), the probability of $u_i$ to visit location $l_j$ must also include multi-variate temporal correlation between the user and the locations that are being appraised to be recommended. Such correspondence is two-fold that have been taken into account in Eq. \ref{Step_2}. From one side, it should reflect the extent of shared temporal activity among user and locations ($\Psi(u_i,l_j)$) where $u_i^s$ and $l_j^s$ are the set of multi-aspect temporal slabs associated with $u_i$ and $l_j$, respectively. From another perspective, the depth of temporal visibility pattern between $u_i$ and $l_j$ must be assessed. Initially, the probability for a $u_i$ to visit $l_j$ is proportionate to the sum of joint probabilities which include temporal latent factors. Nonetheless, we compute the temporal impact through the average value of the joint probability for $u_i$ to visit $l_j$ involving  time-related latent features denoted as ($\sum\limits^{\vbox to 0pt{\hbox{\,\rule{.5pt}{1.35em}}}}_{z_d} \sum\limits^{\vbox to 0pt{\hbox{\,\rule{.5pt}{1.35em}}}}_{z_h} Pr(u_i,l_j,z_h,z_d)$). For clarity purposes, we have modeled two latent factors ($Z=\{z_h,z_d\}$), however, we will generalize the joint probability to incorporate multiple parameters later in this section.
\begin{equation}
\small
\label{Step_2}
Pr(u_i,l_j) \propto \  \  \phi_t \  \Psi(u_i,l_j) +  (1-\phi_t) \sum\limits^{\vbox to 0pt{\hbox{\,\rule{.5pt}{1.8em}}}}_{z_d} \sum\limits^{\vbox to 0pt{\hbox{\,\rule{.5pt}{1.8em}}}}_{z_h} Pr(u_i,l_j,z_h,z_d)
\end{equation}
Let's firstly memorize that we have already exploited multi-aspect temporal slabs as unfolded priorly (Section \ref{Exploiting_Temporal_Blocks}).
\begin{equation}
\label{Step_2_a}
\Psi(u_i,l_j)= \frac{\left | u_i^s \bigcap l_j^s \right |}{\left | u_i^s \bigcup l_j^s \right |}
\end{equation}
As formulated in Eq. \ref{Step_2}, a set of bi-module impacts are tuned in a mixture model using $\phi_t$ parameter, ranging within[0,1]. We have included the tuning process in section \ref{Parameter_Settings}. In addition, as the scales concerning $\Psi(u_i,l_j)$ and $\sum\limits^{\vbox to 0pt{\hbox{\,\rule{.5pt}{1.35em}}}}_{z_d} \sum\limits^{\vbox to 0pt{\hbox{\,\rule{.5pt}{1.35em}}}}_{z_h} Pr(u_i,l_j,z_h,z_d)$ differ, we normalized the values via dividing each probability by the maximum value proposed for every query user. This method is better than feature scaling (\cite{Tax2000}), because the minimum value will not be converted to zero, as of having the null value in recommendation.\\
Similar to \cite{Yuan2013a}, we don't treat user and locations independently conditioned on temporal latent variables. Also, the parameters should be formulated following the TSP feature. Due to the fact that $z_h$ is a subset of $z_d$ (i.e. hour of a day is contained by day of the week) as Eq. \ref{Step_2} denotes, $z_h$ is mentioned before $z_d$ in the joint probability and is placed in the inner loop of the average summation (i.e. $\sum\limits^{\vbox to 0pt{\hbox{\,\rule{.5pt}{1.35em}}}}_{z_d} \sum\limits^{\vbox to 0pt{\hbox{\,\rule{.5pt}{1.35em}}}}_{z_h} \dots $ ).\\
Logically, we now need to explain Eq. \ref{Step_3} which represents the joint probability of user $u_i$ to visit $l_j$ constrained by twin temporal latent parameters of $z_h$ and $z_d$:
\begin{equation}
\small
\label{Step_3}
Pr(u_i,l_j,z_h,z_d)\propto Pr(u_i) Pr_{\nu}(l_j|u_i) Pr(z_h|z_d,u_i,l_j) Pr(z_d|u_i,l_j)
\end{equation}
Primarily, $Pr(u_i)$ is assigned by one, because all the users are treated equally. Also, $Pr_{\nu}(l_j|u_i)$ comprises total non-temporal influences for $u_i$ to visit $l_j$. It is also worth to mention, that if the resulting value regarding multiplication of probabilities is less than the decimal minimum, the joint probability will be ignored as having the value of zero. Hence, in implementation, we take the log from both side of this equation which converts multiplication to summation and prevents missing values to be imputed incorrectly. 
\begin{proof}
	\textit{We can prove Eq. \ref{Step_3} through applying Bayes theorem to $ Pr(z_h,z_d|u_i,l_j)$. As illustrated in Fig. \ref{fig:Latent_Diagram}, users and locations form distinctive graphs while connecting to initial influences ($\nu_{i,j}$) and temporal preferences (here assumed as $z_h$, $z_d$). Therefore we can reorder $Pr(u_i,l_j,z_h,z_d)$ as follows:
	}\begin{equation}
	\small
	\label{Step_4}
	Pr(u_i,l_j,z_h,z_d) \propto Pr(u_i,l_j) Pr(z_h,z_d|u_i,l_j)
	\end{equation}
	\begin{equation}
	\small
	\label{Step_5}
	Pr(u_i,l_j) \propto Pr(u_i) Pr_{\nu}(l_j|u_i)
	\end{equation}
\textit{$Pr_{\nu}(l_j|u_i)$ is proportionate to the joint probability of $ u_i$ and $l_j$. Also, Eq. \ref{Step_6} paraphrases $Pr(z_h,z_d|u_i,l_j)$ through the joint probability of $z_h,z_d$ and rendering $u_i,l_j$ as a single parameter:}
	\begin{equation}
	\small
	\label{Step_6}
	Pr(z_h,z_d|u_i,l_j)=Pr(z_h|z_d,u_i,l_j)Pr(z_d|u_i,l_j)
	\end{equation}		
\textit{By substituting equation \ref{Step_6} and \ref{Step_5} into \ref{Step_4}, we can reach Eq.\ref{Step_3}.}\\
	\end{proof}
    Now that we have formulated both partitions of Eq. \ref{Step_2}, we need to consider that the data is incomplete due to two reasons. Firstly, LBSN users mostly perform a limited number of check-ins and the information regarding their visiting behavior on POIs in various times is insufficient. Secondly, our evaluation method urges data compensations as we exclude certain percentage from one's check-in history and assess how they are retrieved using various methods. Such an embargo makes the data even more defective. Therefore, we propose a model which can attain a user's behavior based on her imperfect visiting log during exploited multi-aspect temporal slabs.\\    
    Apparently, the model owns a set of parameters denoted by $\psi$ including $Pr_{\nu}(l_j|u_i)$, $Pr(z_h|z_d,u_i,l_j)$ and $Pr(z_d|u_i,l_j)$. Here $z_h$ and $z_d$ are the latent variables and $Pr_{\nu}(l_j|u_i)$ can be computed using nontemporal approaches to include other effects. Hence, we aim to maximize the log-likelihood of $\mathcal{L}(\psi)$.\\
	\begin{equation}
	\small
	\label{Step_7}
	\mathcal{L}(\psi)=\sum_{<u_i,l_j> \in <U,L>}^{} log(Pr(u_i,l_j;\psi))
	\end{equation}
	We use Expectation-Maximization (EM) to find parameters $\psi$
	that can maximize the log-likelihood of the historical data.
	\begin{itemize}
		\item In the \textbf{E-step}, since there are two latent variables $z_h$ and $z_d$ in MATI, we update their joint expectation $Pr(z_h,z_d|u_i,l_j)$ according to Bayes rule
		as Equation \ref{Step_8}.
		\begin{equation}
		\small
		\label{Step_8}
		Pr(z_h,z_d|u_i,l_j)=\frac{Pr(u_i,l_j,z_h,z_d)}{\sum_{z_d}\sum_{z_h} Pr(u_i,l_j,z_h,z_d)}
		\end{equation}
		\item In the \textbf{M-step}, we find the new $\psi$ that can maximize the log-likelihood
		as follows:
		\begin{equation}
		\small
		\label{Step_9}
		Pr(z_h|z_d,u_i,l_j)=\frac{ Pr(z_h,z_d|u_i,l_j)}{\sum_{z_h^{\prime}} Pr(z_h^{\prime},z_d|u_i,l_j)}
		\end{equation}		
		\begin{equation}
		\small
		\label{Step_10}
		Pr(z_d|u_i,l_j)=\frac{\sum_{z_h} Pr(z_h,z_d|u_i,l_j)}{\sum_{z_d^{\prime}}\sum_{z_h} Pr(z_h,z_d^{\prime}|u_i,l_j)}
		\end{equation}
		
		The value for ${\sum_{z_d^{\prime}}\sum_{z_h} Pr(z_h,z_d^{\prime}|u_i,l_j)}$ is 1. therefore in implementation we have $Pr(z_d|u_i,l_j) \propto \sum_{z_h} Pr(z_h,z_d|u_i,l_j)$.
		
	\end{itemize}
	\begin{proof}
		\title{\textbf{Including infinite number of temporal latent factors}}\\
\textit{Subject to the TSP feature (i.e. $z_1 \subset z_2 \subset z_3 \dots z_{t-1} \subset z_t$), we can \textit{generalize} Eq. \ref{Step_6} to integrate an infinite number of temporal latent factors (i.e. $z_1,z_2,z_3, \ldots ,z_t$) as shown in Eq. \ref{Step_6_g}.}	
	\begin{multline}
	\small
	\label{Step_6_g}
	Pr(z_1,z_2,z_3, \ldots ,z_t|u_i,l_j)=Pr(z_1|z_2,z_3, \ldots ,z_t,u_i,l_j)  Pr(z_2|z_3,z_4, \ldots ,z_t,u_i,l_j) \\ \ldots  Pr(z_{t-1}|z_t,u_i,l_j) Pr(z_t|u_i,l_j)
	\end{multline}
	
\textit{By substituting Eq. \ref{Step_6_g} and \ref{Step_5} into Eq. \ref{Step_4_g} which is the multi-aspect version of Eq. \ref{Step_4}, we can reach Eq. \ref{Step_3_g} which includes multiple latent time-related variables proposed as the generalized version of the bi-variate model proposed in Eq. \ref{Step_3}.}
\end{proof}
	\begin{equation}
	\small
	\label{Step_4_g}
	Pr(u_i,l_j,z_1,z_2,z_3, \ldots ,z_t) \propto Pr(u_i,l_j) Pr(z_1,z_2,z_3, \ldots ,z_t|u_i,l_j)
	\end{equation}	
	\begin{multline}
	\label{Step_3_g}
	\small
	Pr(u_i,l_j,z_1,z_2,z_3, \ldots ,z_t) \propto Pr(u_i) Pr_{\nu}(l_j|u_i) Pr(z_1|z_2,z_3, \ldots ,z_t,u_i,l_j)\\ Pr(z_2|z_3,z_4, \ldots ,z_t,u_i,l_j) \ldots Pr(z_{t-1}|z_t,u_i,l_j) Pr(z_t|u_i,l_j)
	\end{multline}
\subsection{Hybrid Decision Method}
\label{Hybrid}
Our recommendation system relies on multivariate time-related latent factors. However, such a temporally guided mechanism is supposed to be applicable in all scenarios. We can imagine a two-fold storyline: ($i$) \textit{Cold start case}: When a location has been visited for a limited number of times or a user owns a small number of visits in her check-in log, the temporal influence will neither gain an adequate weight to distinguish time-related metrics for the location nor for the user. In other words, when an LBSN user has registered very few check-ins in various temporal slabs, we cannot strongly predict her time-related mobility pattern.
($ii$) \textit{Erratic mobility pattern}: At times recommender systems can embrace certain users whose temporal behaviors are inconsistent with dataset features. In our work, we learn the joint temporal and mobility pattern observed in the dataset to leverage multi-aspect temporal slabs. However, not necessarily all the users would follow such influences. For instance, owing to holidays a user may go to a restaurant on Monday at 10am and go to a bar afterward. Such a behavior has the least probability for most of the people if their check-in history is business-oriented. This affirms a meaningless temporal correlation between a user and LBSN venues.\\
Due to aforementioned binary scenarios, not all the users can be treated the same through the MATI approach. Hence, we transform our system to a hybrid one that can initially recognize whether each query user is temporally sensitive or not. Subsequently, it can propose proper unvisited POIs based on the time-related weights.\\
As denoted in Eq. \ref{Step_2_a}, $\Psi(u_i,l_j)$ can mimic how a query user $u_i$ and a sample location $l_j$ share temporal check-in activities via jacquard coefficient. On the other hand, $\Psi(u_i,l_j)$ can indicate the extent of temporal correlation between each user/location pair via considering the exploited temporal slabs. From another perspective, our model also calculates $Pr_{\nu}(l_j|u_i)$ which is the non-temporal version of the CF-based probability for the user $u_i$ to visit each location $l_j$. Anyhow, Rather than surveying the temporal correlation between a user and any of the POIs in her check-in log, we opt for the average value of $\Psi(u_i,l_j)$ computed for the query user $u_i$ and any of the recommended locations disregarding the temporal effects. On this way, we can comprehend whether each user, temporally synchronizes with the list of recommended POIs before implanting the temporal effects. If the computed average metric passes the threshold or is fallen in the properly trusted range, we can decide to apply our multi-aspect time-related influence. Otherwise, we can simply exclude temporal features and utilize the primary metrics alongside other non-temporal attributes. Nonetheless, we would need to evaluate our system to see under what condition and constraints we can get the best performance. Hence, via a study on pilot set explained in section \ref{Parameter_Settings} we can make the final decision to choose the best range regarding the average $\Psi(u_i,l_j)$ where we can achieve the best outcome. Finally, we can rely on the threshold to accomplish our hybrid decision method.
\section{Experimental Evaluation}
\label{evaluation}
In this section, through releasing multiple experiments, we compare our proposed method with various competitors, explained in section \ref{Recommendation_Methods}. Firstly, we assure under what tuning metrics, our proposed temporal hybrid framework  will achieve its best performance. Secondly, we study how the Multi-aspect time-related perception can improve baseline methods which merely rely on non-temporal or uni-aspect temporal effects. Nevertheless, we need to take a point into consideration that effectiveness of POI recommendation systems on LBSN datasets are always affected by the low density of User-POI and User-time-POI matrices. However, considering the results, we demonstrate that while MATI model can be utilized to reveal the comprehensive temporal correlation in a general user-item relationship, it can also promote the location recommendation systems in the LBSN sphere.
\subsection{Dataset}
\label{Data_Set}
\begin{figure}[!htp]	
	\tiny
	\center
	\minipage{0.71\textwidth}
	\centering
	\includegraphics[width=\linewidth]{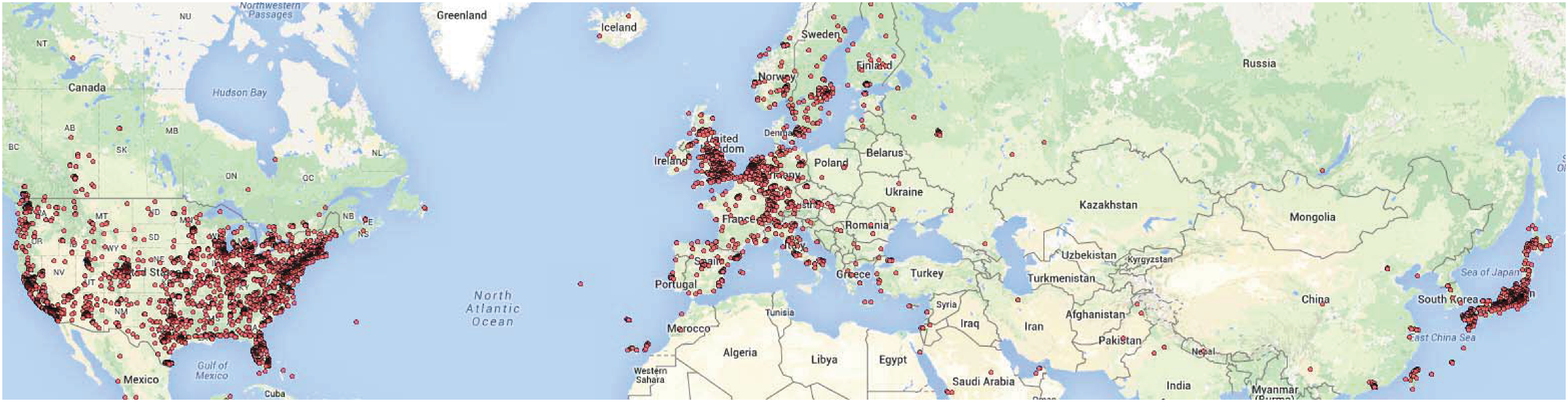} 
	\small (a) Brightkite
	\endminipage\hfill
	\minipage{0.20\textwidth}
	\centering
	\includegraphics[width=\linewidth]{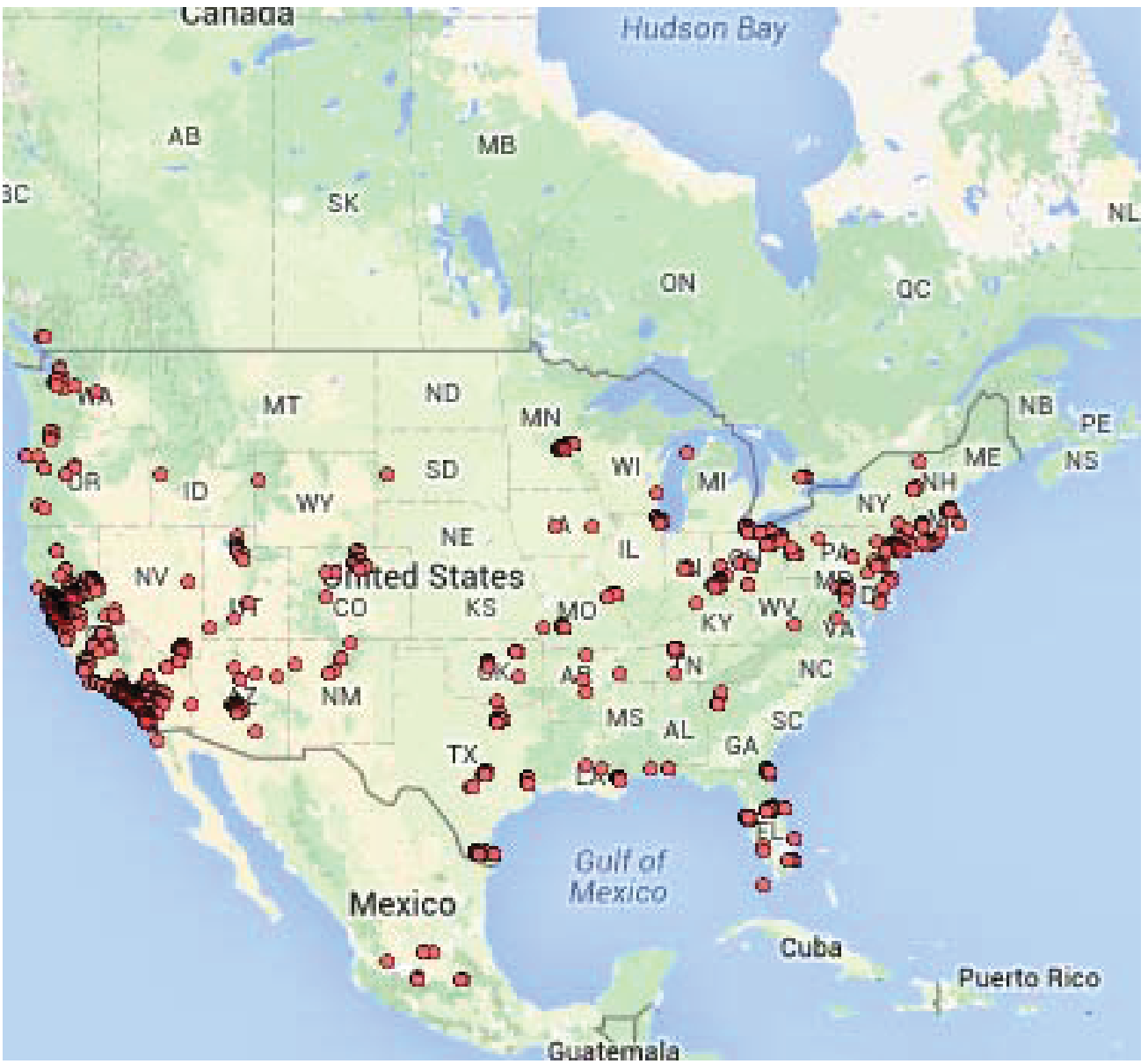} 
	\small (b) Foursquare
	\endminipage
	\caption{\small Check-in Distribution}
	\label{foursquare_brightkite_map}
\end{figure}

Similar to our previous work \cite{Hosseini2016}, we perform the experiments on the same wide-reaching LBSN datasets (\cite{Cho2011}). Both (Foursquare \footnote{http://www.public.asu.edu/~hgao16/} and Brightkite\footnote{https://snap.stanford.edu/data/loc-brightkite.html}) are publicly available. While, the map (Fig. \ref{foursquare_brightkite_map}) shows the spatial distribution of dataset check-ins, pertinent stats are also manifested in table \ref{table_Datasets}. 
\begin{table}[!htp]
	\tiny
	\centering
	\tiny
	\caption{\small Statistics of the datasets}
	\label{table_Datasets}
	\begin{tabular}{lll}
		\hline   & Brightkite & Foursquare \\ 
		\hline 
		Number of users & 58,228 & 4,163 \\
		Number of locations (POIs) & 772,967 & 121,142 \\
		Number of check-ins & 4,491,143 & 483,813 \\
		Number of social links & 214,078 & 32,512 \\
		Cold start ratio (less than 5 POIs) & 53.36\% & 14.17\% \\
		Avg. visited POIs per user & 20.93 & 64.66 \\
		User-POI matrix density& $2.7 \times 10^{-5}$ & $5.33 \times 10^{-4}$ \\
		\hline \end{tabular}
\end{table}
Simply, we can find a high volume of cold start users in Brightkite dataset which comprises more than 50\% of the dataset check-ins. We can witness that the data is scattered (Density: $2.7 \times 10^{-5}$).
In addition, we observe that owing to the fact that merely 8\% of twin users share a minimum of 5 locations, the Foursquare dataset suffers from scarcity which is a common practice in all LBSN data sources.
\subsection{Evaluation Metrics}
\label{Evaluation_Metrics_Section}
Presumably, a POI recommendation task returns top $N$ (i.e. 5, 10 and 20) highly ranked locations for each query user. Two methods can evaluate the effectiveness: (i) The survey-based \textit{normalized Discounted Cumulative Gain}(nDCG) (\cite{Manning2008}) and (ii) F1-score ratios (\cite{Ye2011,Cheng2013a}) which is also used in this paper. Initially, we exclude $x$\% (e.g. 30\%) of the locations from the query user's visiting history. Subsequently, we run the recommendation models using the remaining POIs \footnote{We used \textit{Microsoft SQL Server 2012} relational databases. In expense of the disk space, both non-clustered and clustered indexes which were advised via \textit{Microsoft SQL Server Profiler} accelerated the process speed exceptionally.}. Finally, we count on the truly recovered items. As denoted in Eq. \ref{eq:PrecRecallFmeasure},  considering recommendation@N, The evaluation indicators are  respective total Number of recovered POIs ($R_p$) and the number of initially excluded POIs($E_p$). Precision, Recall, and F1-score metrics are firstly calculated for each query user in test subset (20\% of all dataset users) and final metric is calculated through the total average. $F1-score@N$ will be the final performance balance to find the best among all recommendation models.
\begin{equation}
\small
\label{eq:PrecRecallFmeasure}
\begin{array}
{lcl}
Precision@N=\frac{R_p}{N},
Recall@N=\frac{R_p}{E_p}, 
F1-score@N= \frac {2\times Precision@N \times Recall@N} {Precision@N + Recall@N}
\end{array}
\end{equation}
\subsection{Recommendation Methods}
\label{Recommendation_Methods}
We compare five recommendation methods in the experiments. Among them, the first two are non-temporal and the middle two merely consider one aspect of the time, while the last one is our model which integrates multiple time-aspects. The ultimate aims of the experiments are to signify that our proposed approach outperforms other adversaries as follows.
\begin{itemize}
	    \item \textbf{UBCF}: The primary collaborative filtering method which excludes enhancing influences.
	    \item \textbf{USG}: This method takes advantage of the collaborative filtering method alongside enhancing effects such as social and geographical where $0<\alpha< 1$ and $0<\beta< 1$ (\cite{Ye2011}). UTP-based model (\cite{Yuan2013}) proposes locations at a query time which is a various problem.
	    \item \textbf{USGT}: Uni-aspect Time-related model \cite{Hosseini2016} which is reviewed in section \ref{Temporal_Influence}.
	    \item \textbf{UBCFT}: Another version of \cite{Hosseini2016} which treats all the locations the same in the computation of the user act.    
	    \item \textbf{MATI}:Is the perfect model we proposed in this paper. While prohibiting any conflict with other effects, this method is capable of integrating Multi-Aspect Time-related Influences into CF methods, no matter whether they are model-based or memory-based. We target to substantiate its supremacy.
\end{itemize}
\subsection{Parameter Settings} 
\label{Parameter_Settings}
\begin{figure}[!htp]	
	\tiny
	\center
	\minipage{0.33\textwidth}
	\centering
	\includegraphics[width=\linewidth]{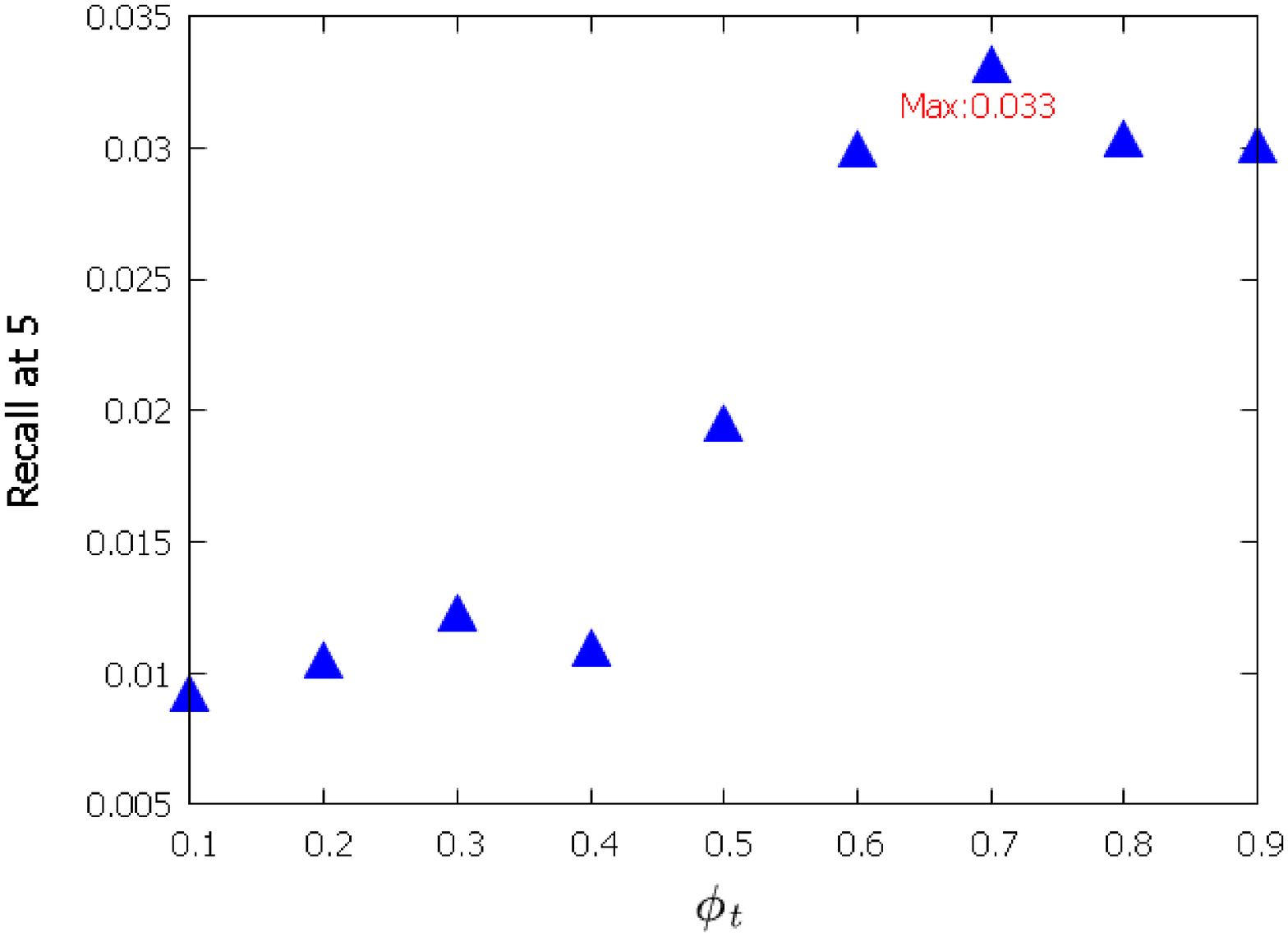} 
	\small (a) Recall
	\endminipage\hfill
	\minipage{0.33\textwidth}
	\centering
	\includegraphics[width=\linewidth]{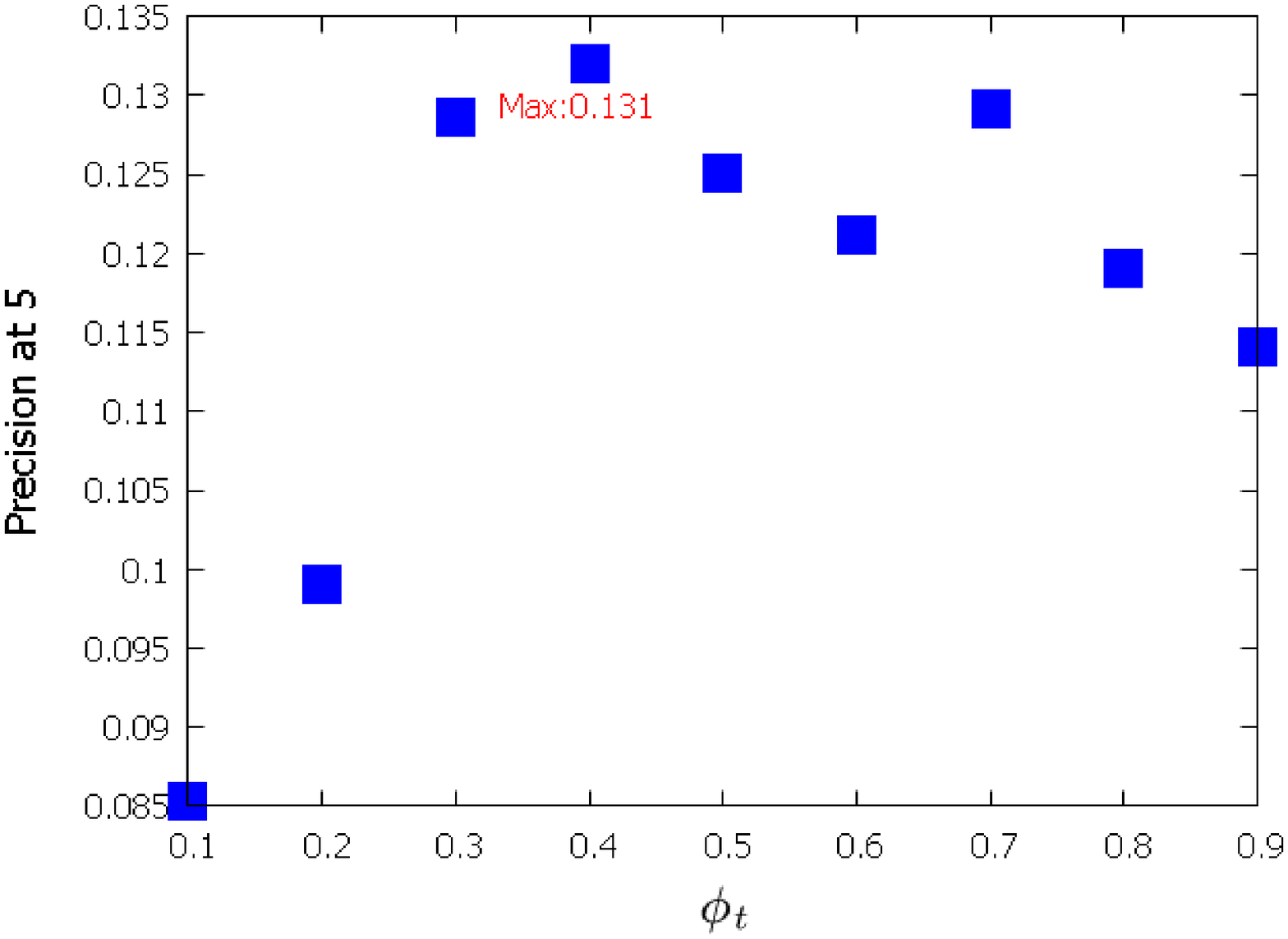} 
	\small (b) Precision
	\endminipage\hfill
	\minipage{0.33\textwidth}
	\centering
	\includegraphics[width=\linewidth]{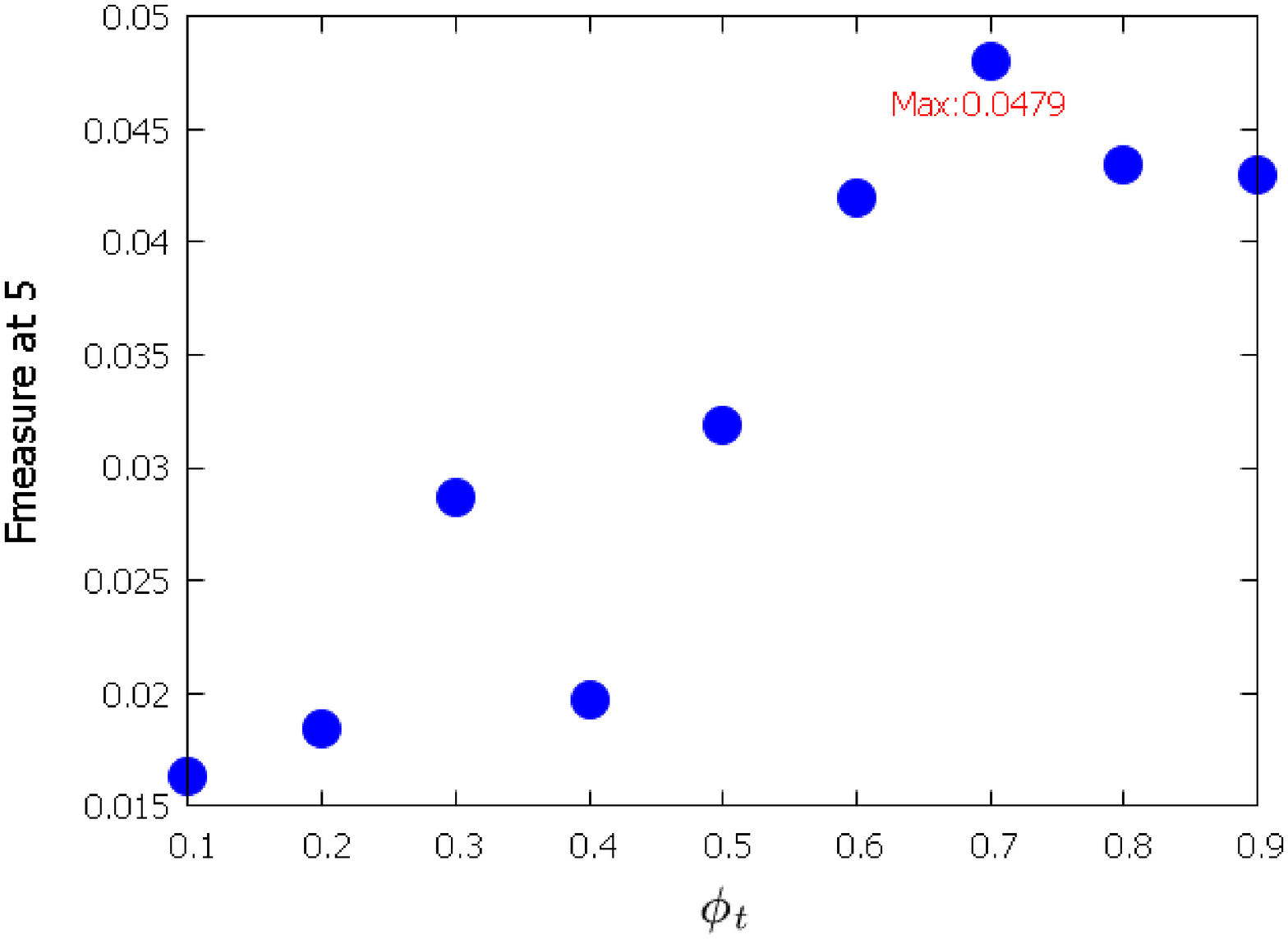} 
	\small (c) F1-Score/Performance
	\endminipage
	\caption{$\phi_t$ at 5 - Brightkite}
	\label{phitat5Brightkite}
\end{figure}

\begin{figure}[!htp]	
	\tiny
	\center
	\minipage{0.33\textwidth}
	\centering
	\includegraphics[width=\linewidth]{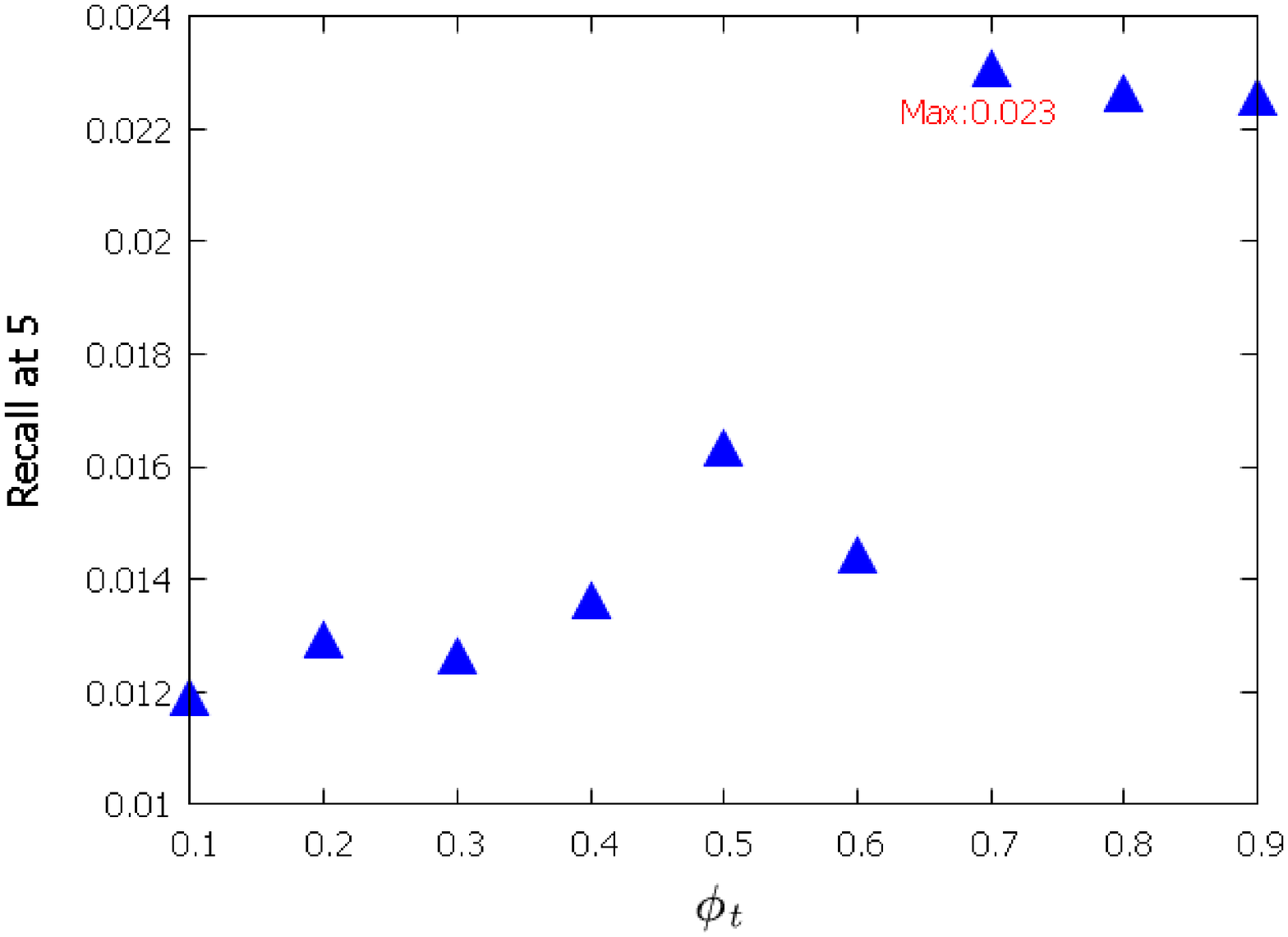} 
	\small (a) Recall
	\endminipage\hfill
	\minipage{0.33\textwidth}
	\centering
	\includegraphics[width=\linewidth]{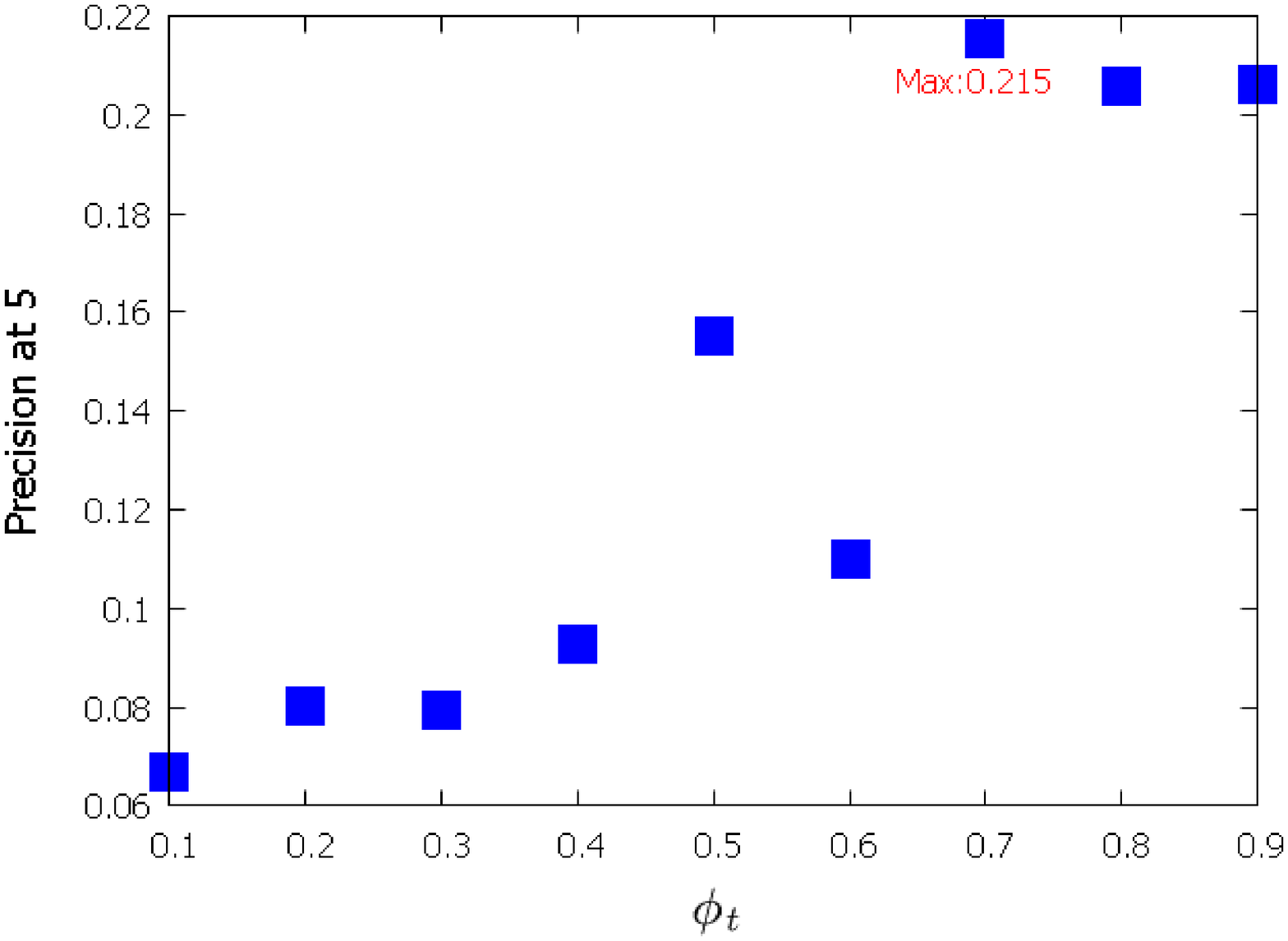} 
	\small (b) Precision
	\endminipage\hfill
	\minipage{0.33\textwidth}
	\centering
	\includegraphics[width=\linewidth]{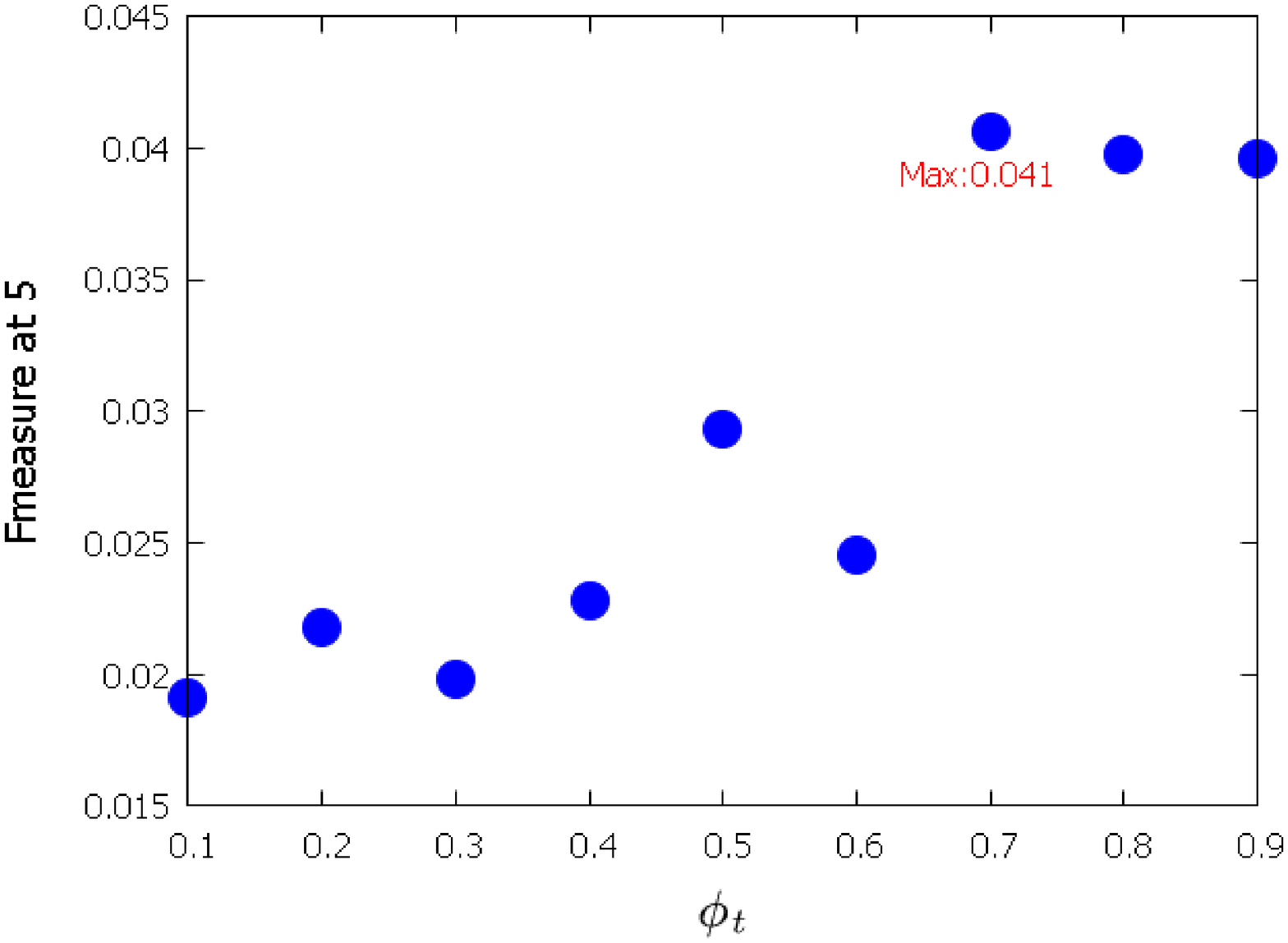} 
	\small (c) F1-Score/Performance
	\endminipage
	\caption{$\phi_t$ at 5 - Foursquare}
	\label{phitat5Foursquare}
\end{figure}

Basically, proposing a recommendation model to comprise multiple aspects of the time is inevitable. Nevertheless, the time factor involves a trade-off process among advantages and defects where parameter settings play a key role in maximizing the effectiveness of recommendation systems. Accordingly, in this section, we explain the way we analyzed temporally influencing parameters for our proposed method (MATI) through a set of tuning experiments. Also, note that we adopted another series of evaluations to assure our method overcomes other competitors based on the performance metrics. Notable parameters to adjust are two-fold. (i) $\phi_t$ (ii) $\sum\limits^{\vbox to 0pt{\hbox{\,\rule{.5pt}{1.35em}}}}_{l_j \in u_i^p} \Psi(u_i,l_j)$\\
It is also worth mentioning that, the visiting histories possessed by cold start users are not reliable to supply adequate evidence concerning multi-aspect temporal slabs. This stems unreal results in spatial item recommendation. For instance, owing to excessive data incompleteness, such users can be associated with a specific temporal slab and pretermit others. Hence, while choosing to analyze the parameters using active and semi-active users (at least 15 check-ins in the log $\left | L_i \right| \geq 15$), we selected a random set of 20\% from both datasets.\\
\textbf{Adjusting $\mathbf{\phi_t}$}: with regard to Eq. \ref{Step_2}, we tuned $\phi_t$ (ranging within [0,1]) as the significant parameter in recommendation mixture model. The essential aim was to figure out the importance coefficiency of the binary factors including the depth of temporal visibility pattern($\sum\limits^{\vbox to 0pt{\hbox{\,\rule{.5pt}{1.35em}}}}_{z_d} \sum\limits^{\vbox to 0pt{\hbox{\,\rule{.5pt}{1.35em}}}}_{z_h} Pr(u_i,l_j,z_h,z_d)$) as well as the extent of shared temporal activity($\Psi(u_i,l_j)$).\\
Figures \ref{phitat5Brightkite} and \ref{phitat5Foursquare} illustrate the parameter tuning results for $\phi_t$ considering the exclusion rate of 30\% while we perform recommendation @5.
In fact, F1-Score measure integrates both precision and recall. Therefore, it is used to make the final decision. When the value for $\phi_t$ is set to 0.7 in both datasets we achieve the best performance. This affirms that the extent of shared temporal activity between every test user and the proposed POI is more important than the depth of temporal visibility pattern. On the other hand, as higher the number of shared temporal slabs, bigger is the level of temporal influence.\\
During tuning, we firstly fixed the value of $\phi_t$ for the mixture model. Subsequently, we continued by exploiting the best range in which we could make the proper decision for the hybrid recommendation system.\\
\textbf{Adjusting $\mathbf{\sum\limits^{\vbox to 0pt{\hbox{\,\rule{.5pt}{1.35em}}}}_{l_j \in u_i^p} \Psi(u_i,l_j)}$:} From another perspective, we explored the best decimal range (e.g. [0.4-0.9]) in which the average value for the shared temporal activity could reach the highest performance in the recommendation. Here, $u_i$ is the query user who is proposed with the set of locations denoted as $u_i^p$. The selected range can maximize the performance of our proposed hybrid framework (discussed in section \ref{Hybrid}). On the other hand, if the computed average metric for a user is between the selected range, the hybrid system can enforce multi-aspect temporal influence. Otherwise, while debarring the time-related latent factors, it can merely rectify the non-temporal method (i.e. CF based plus non-temporal effects).\\
\begin{figure}[!htp]	
	\tiny
	\center
	\minipage{0.49\textwidth}
	\centering
	\includegraphics[width=\linewidth]{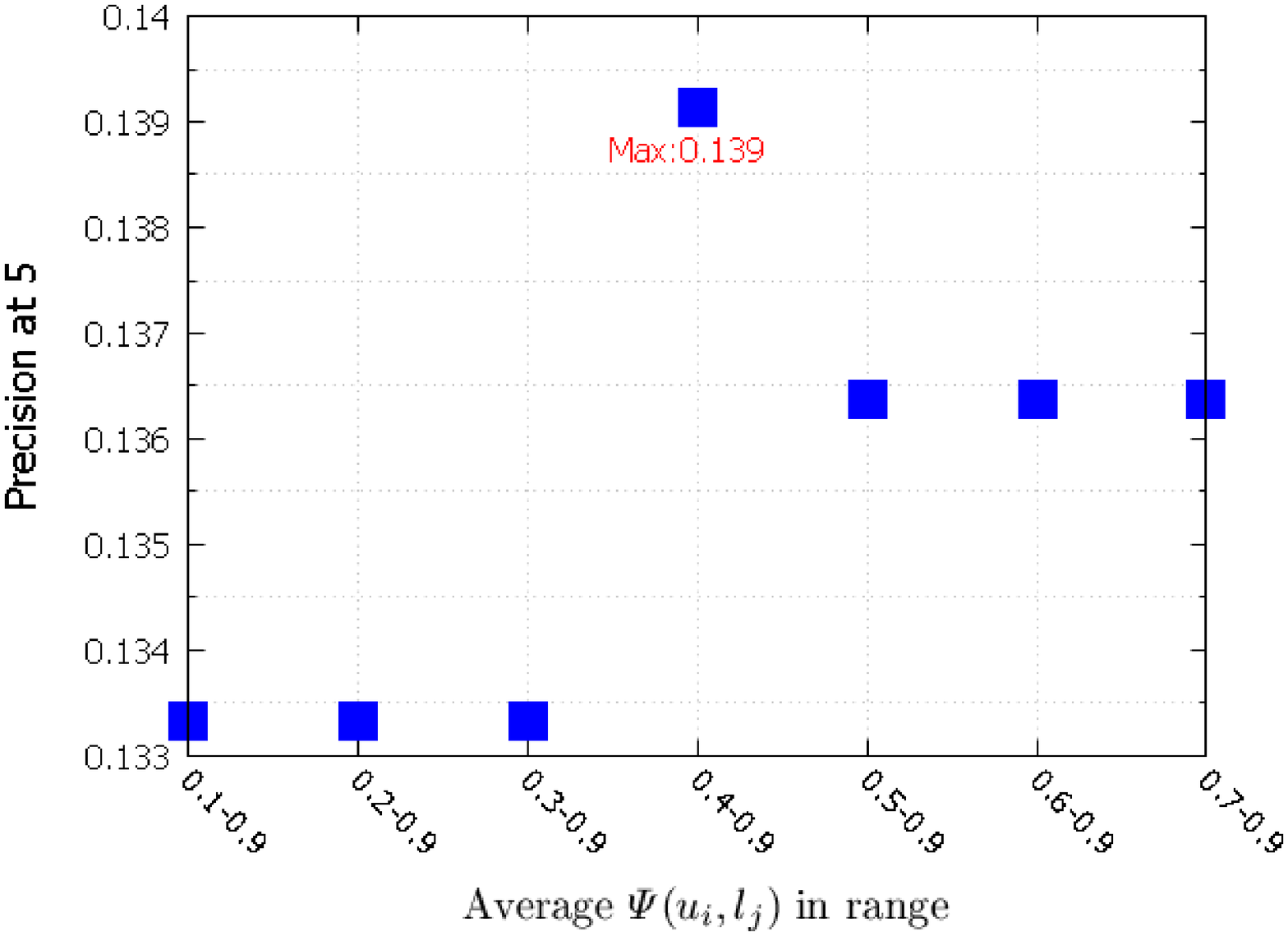} 
	\small (b) Precision
	\endminipage\hfill
	\minipage{0.49\textwidth}
	\centering
	\includegraphics[width=\linewidth]{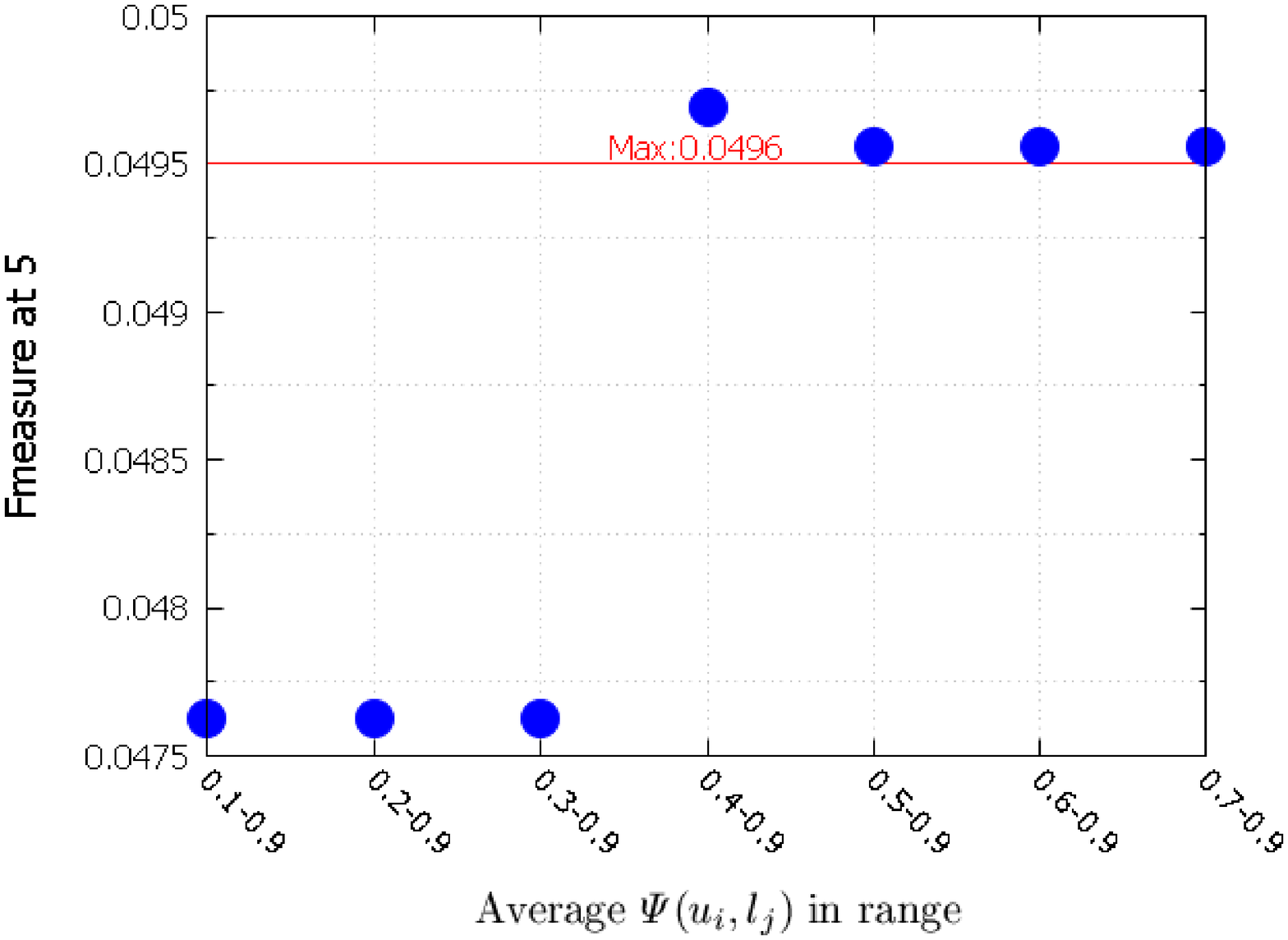} 
	\small (c) Performance
	\endminipage
	\caption{Average $\Psi(u_i,l_j)$ at 5- Brightkite}
	\label{AveragePsiat5Brightkite}
\end{figure}

\begin{figure}[!htp]	
	\tiny
	\center
	\minipage{0.49\textwidth}
	\centering
	\includegraphics[width=\linewidth]{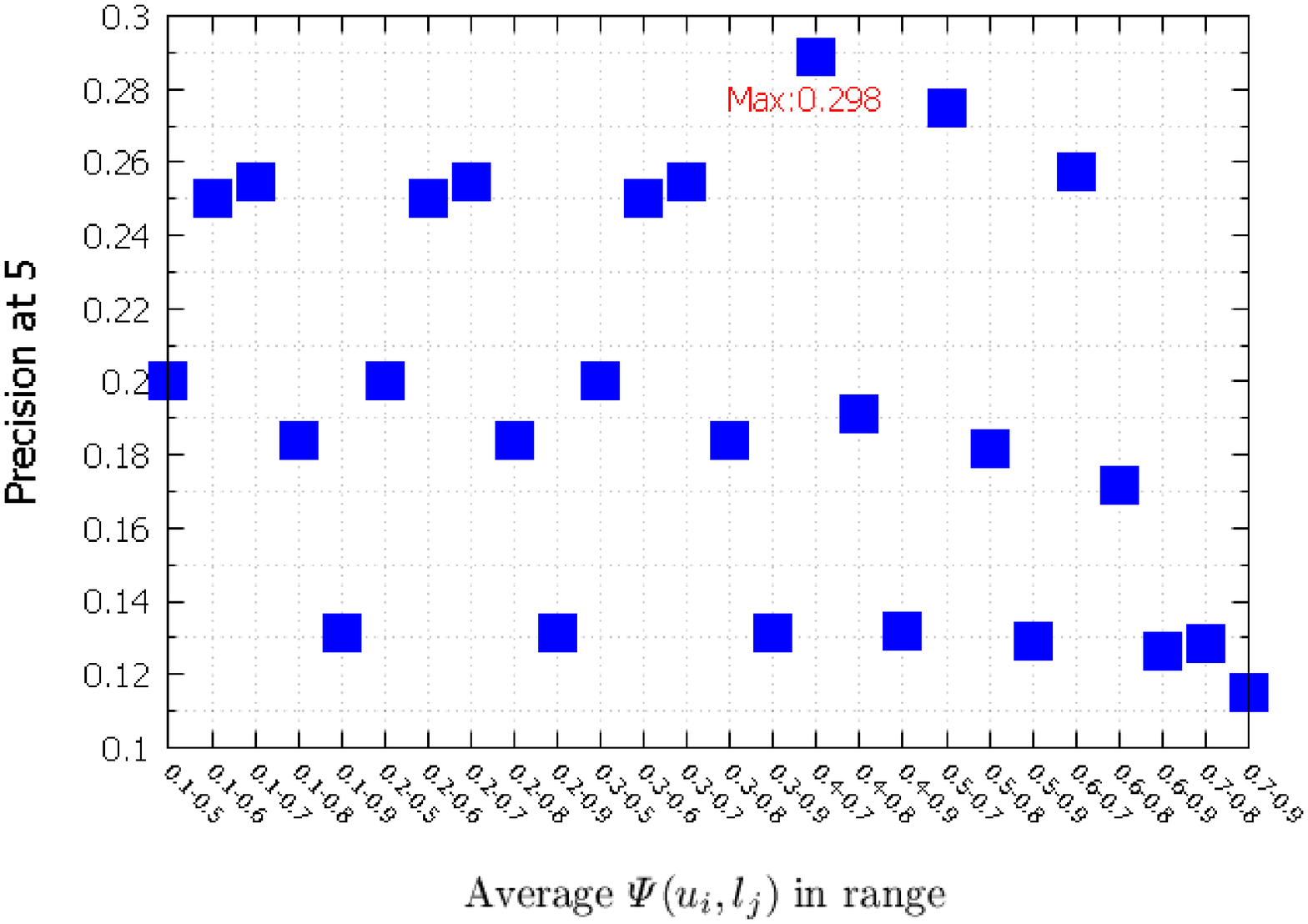} 
	\small (b) Precision
	\endminipage\hfill
	\minipage{0.49\textwidth}
	\centering
	\includegraphics[width=\linewidth]{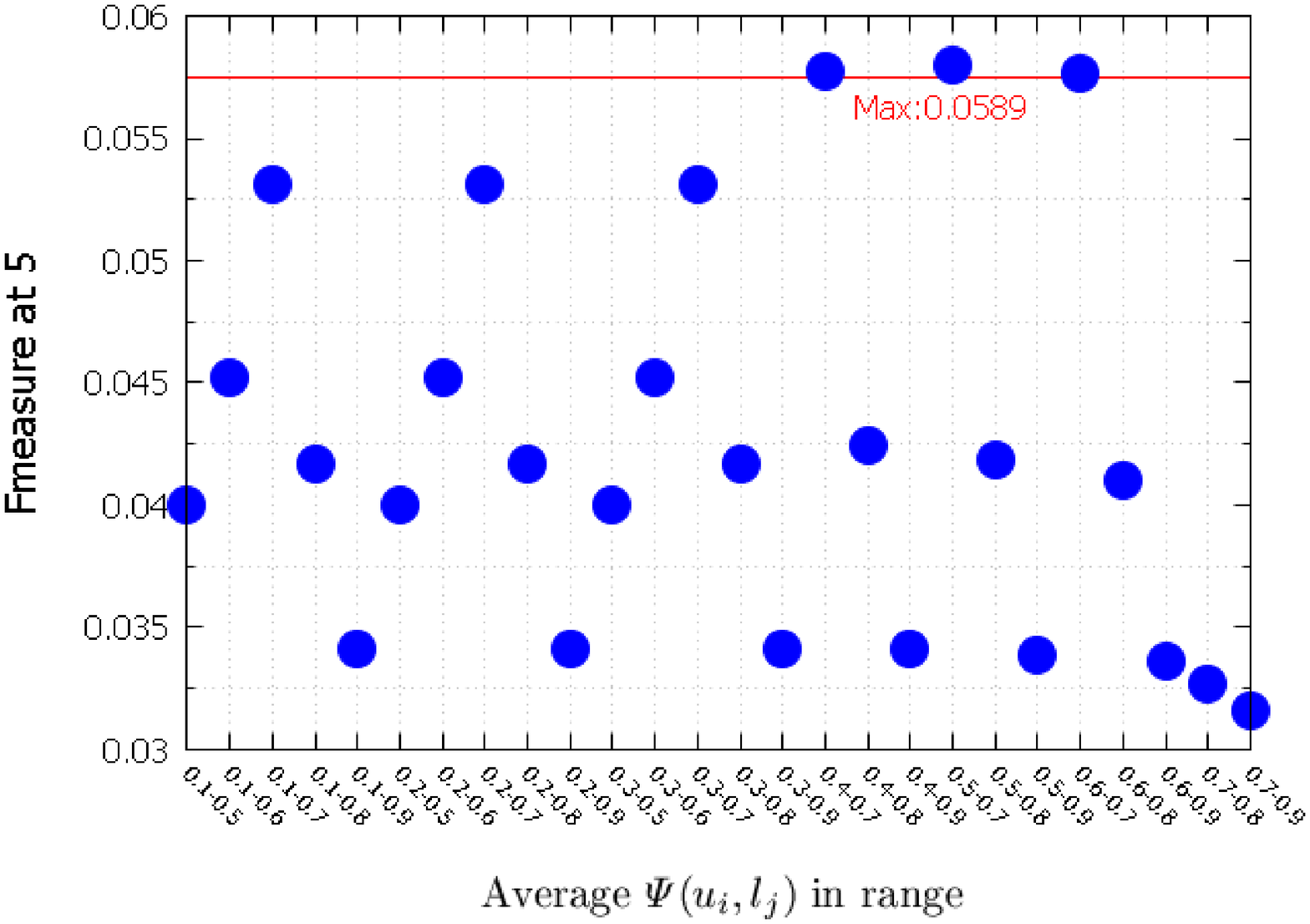} 
	\small (c) Performance
	\endminipage
	\caption{Average $\Psi(u_i,l_j)$ at 5- Foursquare}
	\label{AveragePsiat5Foursquare}
\end{figure}
As shown in Fig. \ref{AveragePsiat5Brightkite} and Fig. \ref{AveragePsiat5Foursquare}, the best ranges in which we can apply the temporal influence are [0.4-0.9] and [0.4-0.8] (Union of the two best ranges of [0.4-0.7] $\cup$ [0.6-0.8]) for respective Brightkite and Foursquare datasets.
\begin{table}[!htp]
	\centering
	\caption{\small USG Optimized values}
	\label{USG_Optimised_values}
	\begin{tabular}{|l|l|c|}
		\hline
		\multirow{2}{*}{}    & \multicolumn{2}{l|}{{\bf F1-Score @5}}                     \\ \cline{2-3} 
		& \multicolumn{1}{c|}{{\bf $\alpha$}}              & \multicolumn{1}{c|}{{\bf $\beta$}} \\ \cline{2-3}  \hline
		{\bf  Foursquare } & \multicolumn{1}{c|}{0.2} & \multicolumn{1}{c|}{0.6}\\ \hline
		{\bf  Brightkite } & \multicolumn{1}{c|}{0.3} & \multicolumn{1}{c|}{0.4}\\ \hline
	\end{tabular}
\end{table}
Moreover, we provide some information about parameter settings of the baseline models. For USGT, with regard to Eq. \ref{eq:pijlambda}, we assume $\lambda=0.5$ to equalize $w_d$ and $w_e$ treatments. Similar to our prior work, we chose a random set of 20\% from users in both datasets and evaluated the rate of neutral POIs ($\{\forall p_y \in \rho | p_y^a=0\}$) among top \textit{K*@Num} proposed locations. owning to the fact that the rate of them was less than 10\%, we set the value for $\xi$ to 0.1 in both datasets. In addition, we set $K$ to 10. Subsequently, in order to set the parameters for USG (\cite{Ye2011}), unlike rank learning (e.g. SVM-pairwise and EM), we iterated the values of $\alpha$ and $\beta$ through 0 and 1, while aiming to obtain the best performance @5(Table \ref{USG_Optimised_values}). Finally, the best parameters of $\alpha$ and $\beta$ were selected based on the highest F1-score@5.\\
\subsection{Performance Comparison}
In this subsection, we describe the comparison outcomes among our proposed model and other
competitive rivals via utilizing well-tuned parameters. While theoretically, it is appealing to propose a model which can simultaneously comprise multiple temporal scales in a recommendation task, we additionally report the evaluation results to assure how our novel model excel others.\\
\begin{figure}[!htp]
	\tiny
	\center
	\minipage{0.33\textwidth}
	\centering
	\includegraphics[width=\linewidth]{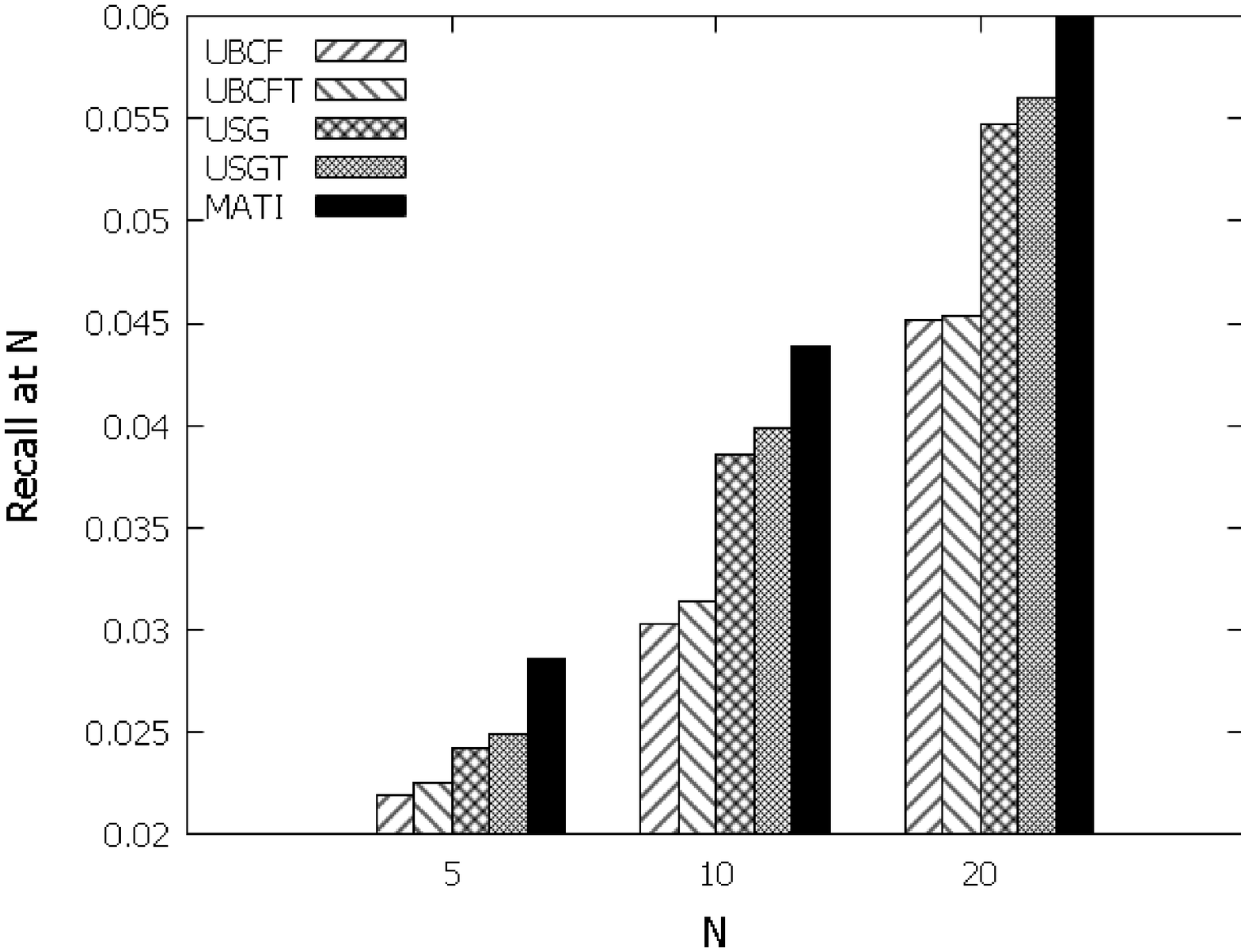} 
	\small (a) Recall
	\endminipage\hfill
	\minipage{0.33\textwidth}
	\centering
	\includegraphics[width=\linewidth]{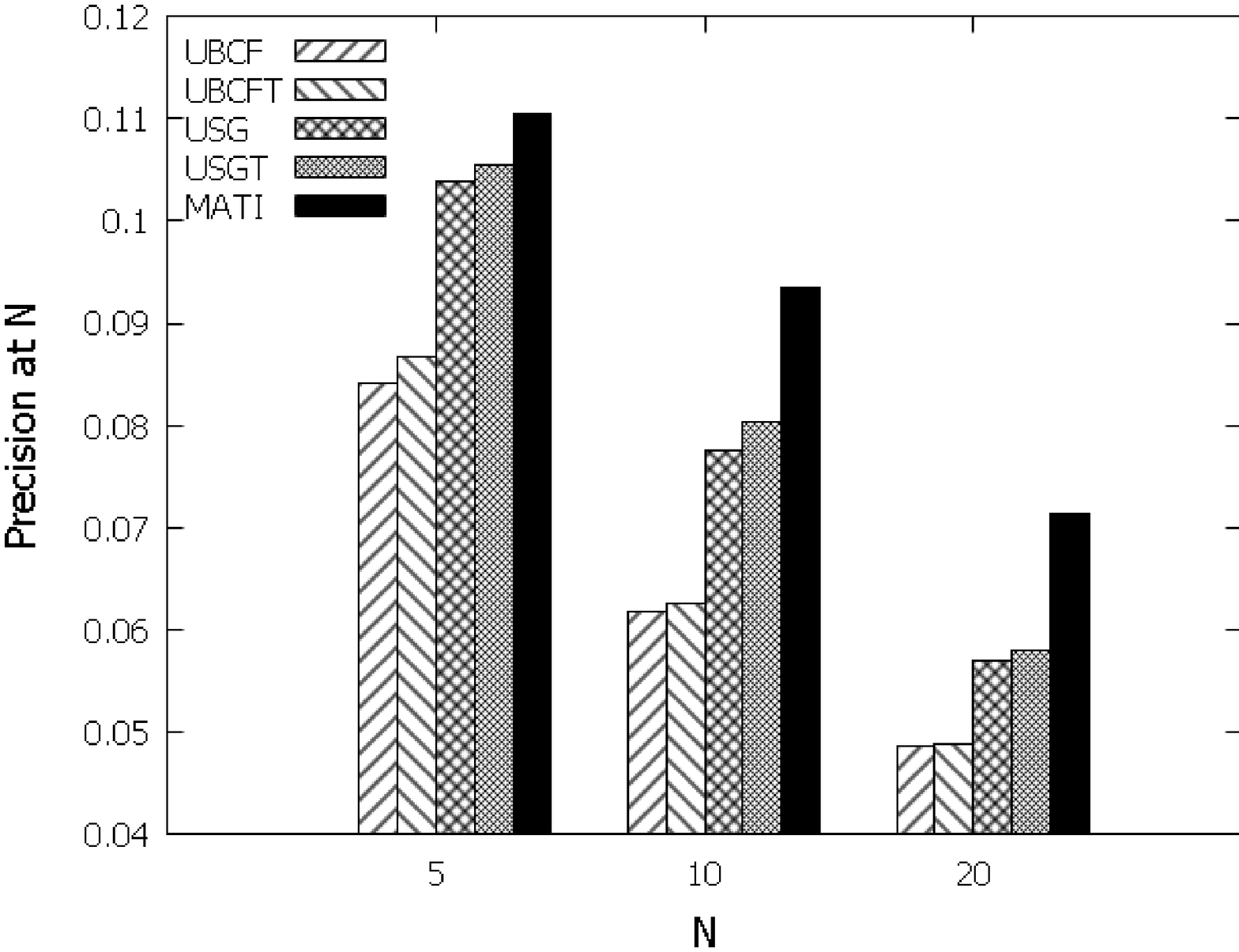} 
	\small (b) Precision
	\endminipage\hfill
	\minipage{0.33\textwidth}
	\centering
	\includegraphics[width=\linewidth]{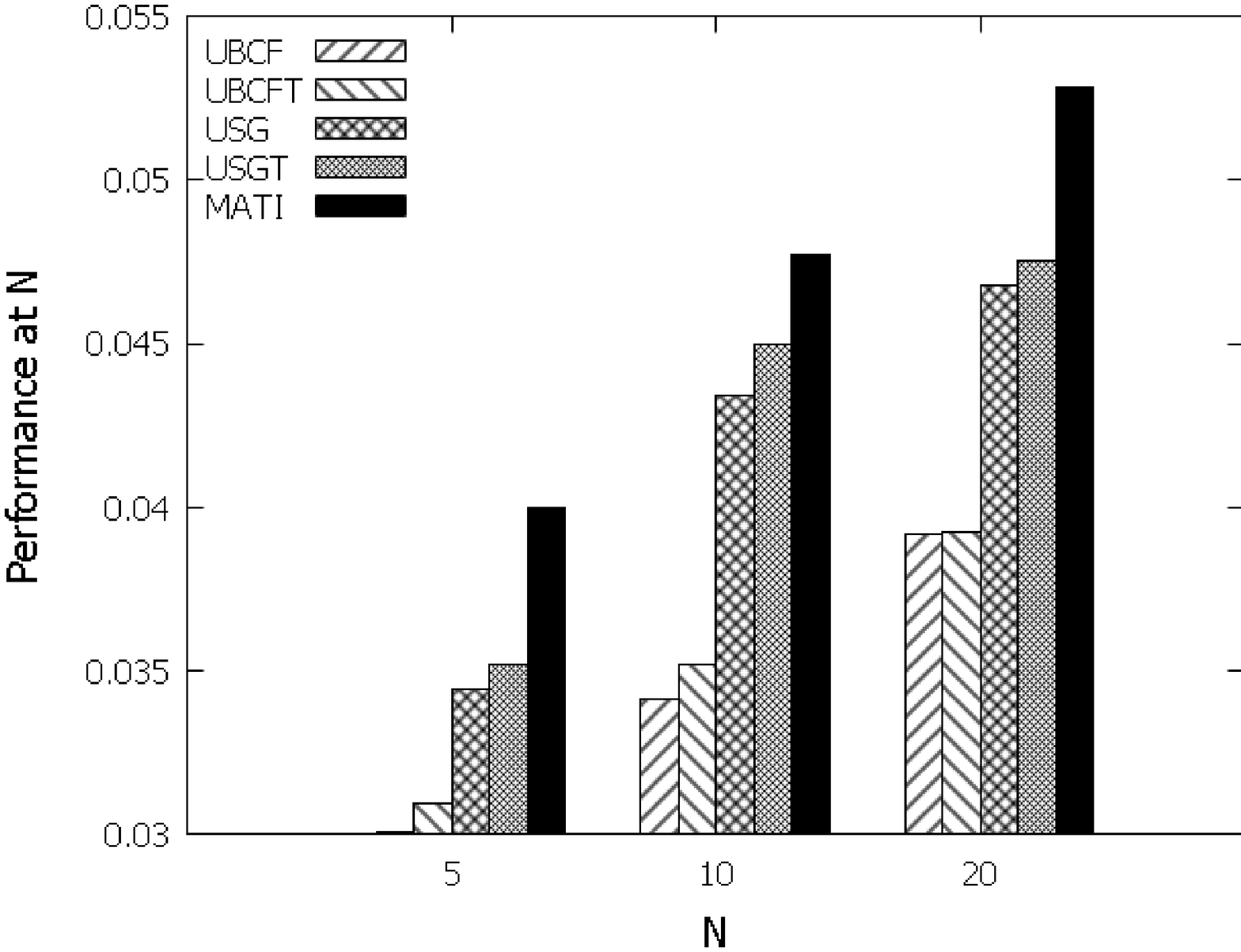} 
	\small (c) F1-Score/Performance
	\endminipage
	\caption{\small Comparing the methods - Foursquare dataset}
	\label{finalparameters_fs}
\end{figure}
\begin{figure}[!htp]	
	\tiny
	\center
	\minipage{0.33\textwidth}
	\centering
	\includegraphics[width=\linewidth]{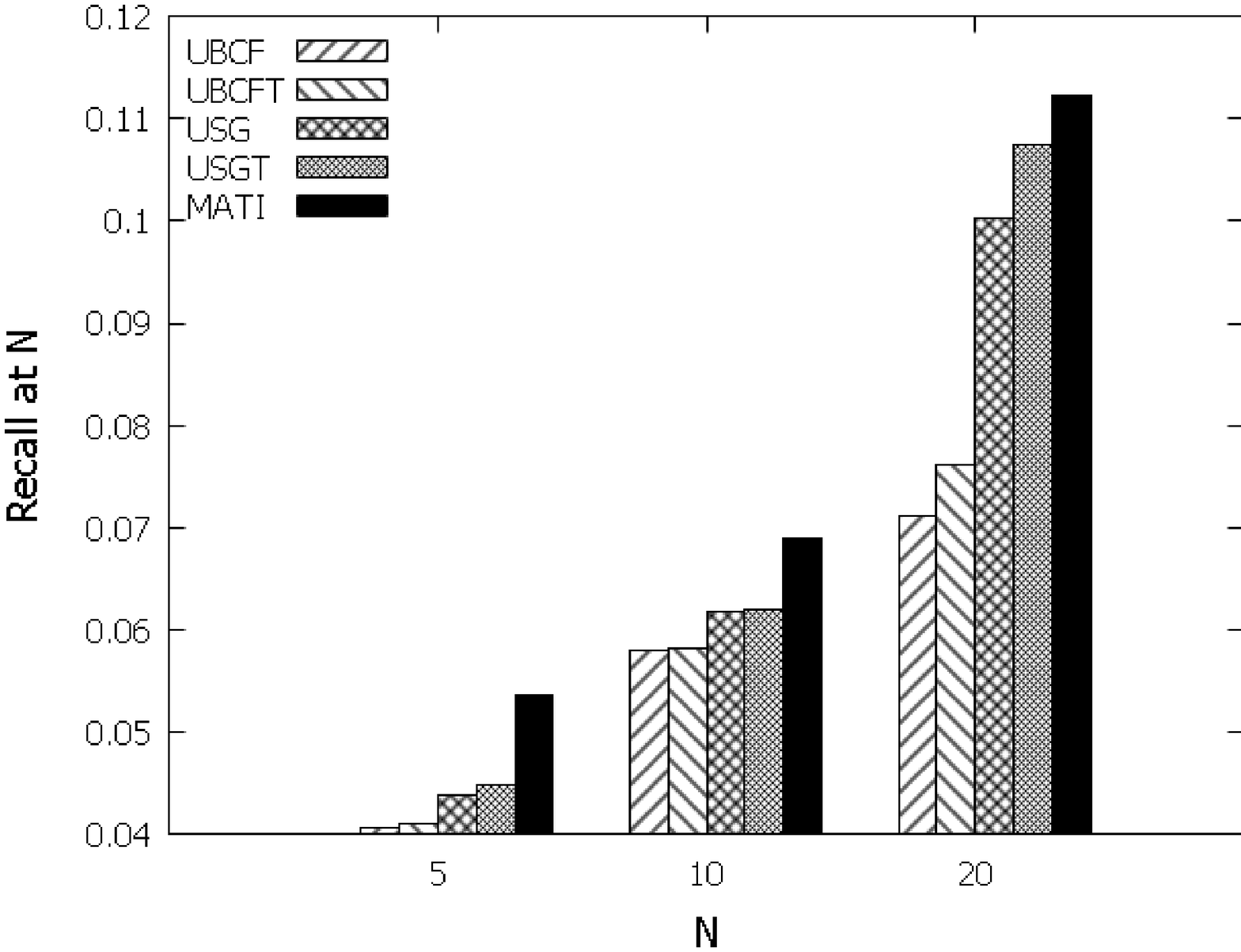} 
	\small (a) Recall
	\endminipage\hfill
	\minipage{0.33\textwidth}
	\centering
	\includegraphics[width=\linewidth]{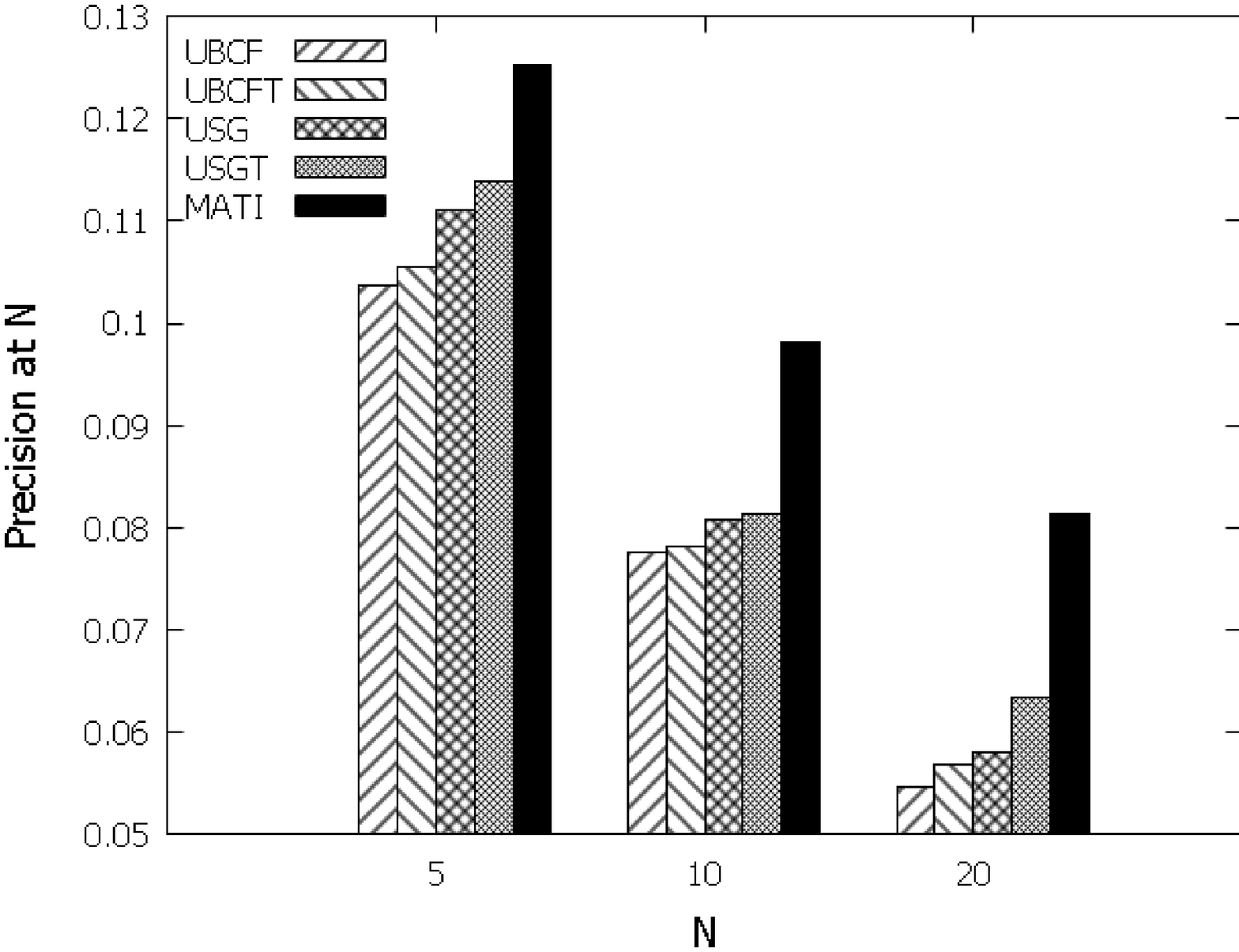} 
	\small (b) Precision
	\endminipage\hfill
	\minipage{0.33\textwidth}
	\centering
	\includegraphics[width=\linewidth]{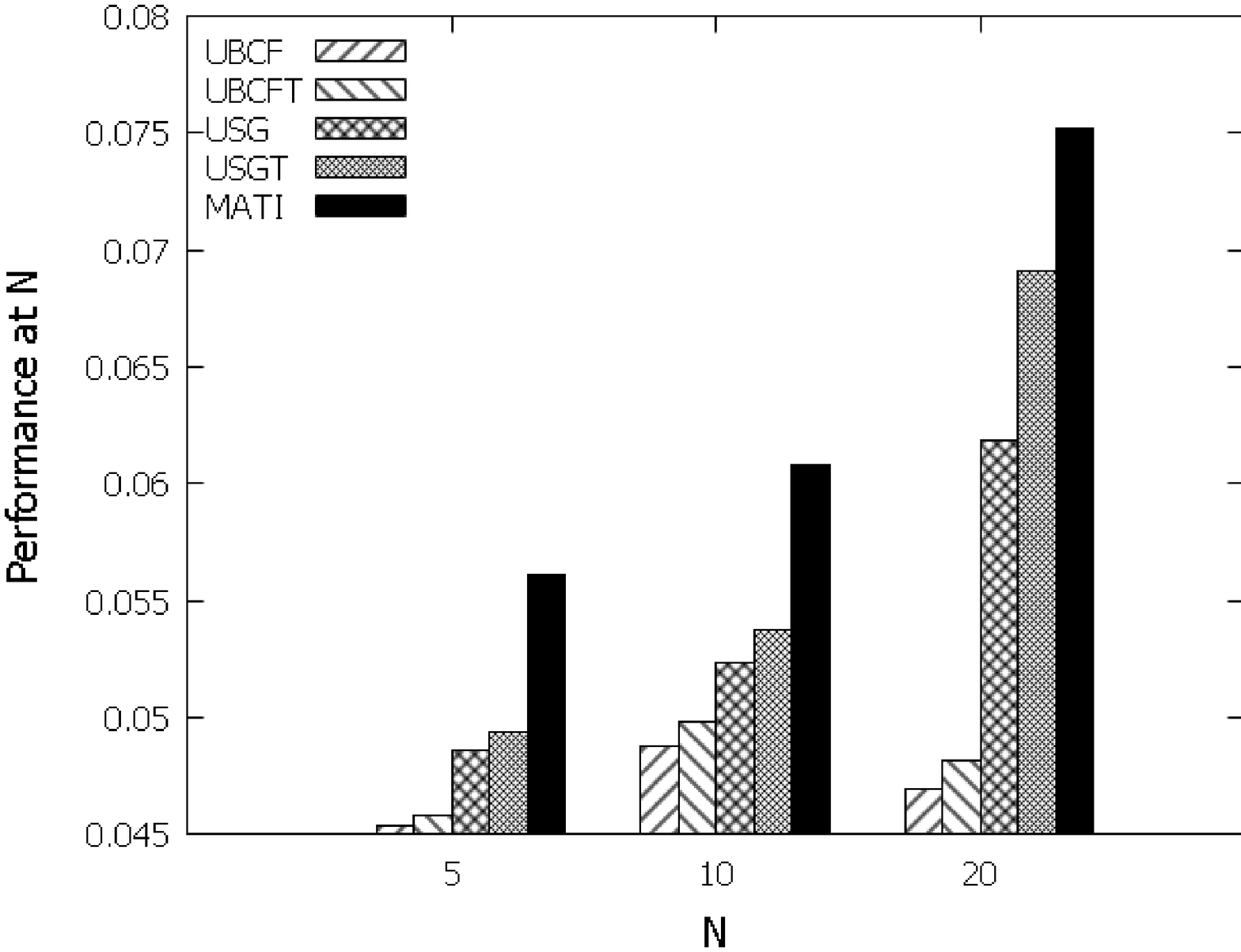} 
	\small (c) F1-Score/Performance
	\endminipage
	\caption{\small Comparing the methods - Brightkite dataset}
	\label{finalparameters_bk}		
\end{figure}
The evaluation aspects are two-fold: (i) Firstly, Figures \ref{finalparameters_fs} and \ref{finalparameters_bk} illustrate respective performance results regarding the recommendation methods on Foursquare and Brightkite datasets. Practically, as a limited number of top-N items are commonly desirable in a recommendation process, we merely compare the performance where N is set to 5, 10, and 20. The figures clearly confirm that our method (MATI) is performing better than other models in terms of top-N recommendations.
(ii) Secondly, we have evaluated how our method can alleviate the rate of failure in proposing one or more true suggestions  (i.e. the POIs which are retrieved after exclusion in pre-processing) to every query user in test pilot set. The number of failures in recommendation@5 is bigger than both recommendations at 10 and 20. Hence, considering recommendation @5, as table \ref{table_RateOfFail} demonstrates, our method has been able to increase the rate of success by 4.1\% and 6.6\% in Brightkite and Foursquare datasets, respectively.\\
\begin{table}[!htp]
	\tiny
	\centering
	\tiny
	\caption{\small Comparing the methods - Rate of failures in recommendation process}
	\label{table_RateOfFail}
	\begin{tabular}{lll}
		\hline  Rate of failures @5 & Brightkite & Foursquare \\ 
		\hline 
		 \textbf{MATI} & 52.3\% & 51.6\% \\
		 \textbf{Best case} of other methods & 55.7\% & 58.2\% \\
		\hline \end{tabular}
\end{table}
Densities concerning User-POI and User-Time-POI matrices are extremely low which is common in LBSN datasets. Therefore, the effectiveness of POI recommendation systems doesn't reach high. For instance, Precision in \cite{Yuan2013} and recall in \cite{Liu2015} are less than 4\% while \cite{Gao2013a} achieve less than 3.5\% for both metrics. Consequently, algorithms are evaluated relatively (i.e. rate of improvement for one versus the baselines).\\
As a matter of fact, where a test user owns a higher number of check-ins in her visiting history, more evidence regarding her temporal activity pattern will be mined and subsequently, our model can better detect temporal correlations between the user and proposed locations. Nevertheless, we observe that the recall rate in all experiments (including tuning) is promoted less than precision. The reason for this occurrence is that the MATI model proposes true recommendations for many active or semi-active users for whom the prior models fail. Accordingly, for a sample test user, precision@5 elevates by 20\% merely due to a single correct proposed POI. But, for Recall@5, as most of such users possess a high number of visits, the value doesn't increase considerably. For instance, if the user has 100 check-ins, then his recall will be improved approximately 0.033 (1 recovered from 30 excluded POIs). Ultimately, as the recall is still a small value, the overall f1-score will not be inflated either.\\
\section{Related Work}
\label{Related_Work}
Nowadays, LBSN platforms (e.g. \textit{Foursquare, Gowalla, Google places, and etc}) shape the essential part of people's daily lives. Accordingly, POI recommendation via such mediums has become a ubiquitous task \cite{Bao2014}. Some traditional methods like \textit{HIT-based}(\cite{Zheng2009}) and \textit{Random Walk \& Restart} (\cite{Tong2006,Xiang2010}) have already been used for location recommendation \cite{Hosseini2016}. Furthermore, various factors \cite{Yin2015} such as \textit{geographical}, \textit{social}, \textit{context-oriented} (e.g. text contents and word-of-mouth) and \textit{temporal} influences have been recently integrated into \textit{Collaborative Filtering} (CF) to improve the performance. In this paper, we have also considered the time as a multi-aspect influential parameter.\\
\textbf{\textsl{Collaborative Filtering (CF).}} While CF-based methods \cite{Ye2011,Levandoski2012,Yuan2013,Yin2015} are dominantly employed in location recommendation systems, they infer the query user's preference regarding every proposing unvisited POI. Collaborative Filtering is categorized into memory-based and model-based approaches \cite{Cheng2013a}. Memory-based approaches have two types of user-based (\cite{Ye2011}) and item-based \cite{Ding2005} which propose unvisited locations to a user based on similarity weights (e.g. \textit{Cosine} and \textit{Pearson} metrics) computed among users and items respectively. Like our prior work \cite{Hosseini2016}, we also utilize user-based collaborative filtering. From data perspective, CF-based methods have been commonly used to perform recommendation task on various data types such as semantics \cite{Hu2013}, Trajectories (\cite{Zheng2010},\cite{Leung2011},\cite{Hung2011}), and check-in logs (\cite{Bao2014}). However, as original CF methods fail to achieve a reasonable performance, other components (e.g. Social and Geographical influence) are jointly amended to enhance recommendation results.\\
\textbf{\textit{Social link}} The correlations among network friends affect any user-item matrix \cite{Ye2012}. In reality, we may visit a POI which has already been visited by a friend on the network. Accordingly, the social links in LBSN sphere \cite{Zhang2015b,Cheng2012,Gao2012} influence users' visibility patterns. Goyal et al. \cite{Goyal2010} study how the social link can affect an individual's decision to visit a location. They also model how the influence is propagated in social networks in a course of time. Ye et al. \cite{Ye2011} study the Jaccard similarity coefficient w.r.t. both locations and friends. However, the parameter settings on their recommendation task confirm that the number of shared locations among two users has a bigger impact on visibility patterns than the number of friends they have in common \cite{Ye2011}. While, Cheng et al. \cite{Cheng2012} similarly claim a minor influence for the social factor in location recommendation, LTSCR \cite{Zhang2015b} and \cite{Cho2011}, concurrently model social data jointly with spatio-temporal information.\\
\textbf{\textit{Geographical Influence (GI).}} Geographical influence has already been studied in several previous works (\cite{Zhang2015c,Ye2011,Wang2012,Yuan2013,Zhang,Liu2013}) and explains why LBSN users tend to visit the POIs which are near to the venues where they have already visited \cite{Hosseini2016}. Such effect has already been modeled using \textit{Power law distribution} (\cite{Ye2011,Wang2012,Yuan2013}) and \textit{Multi-Center Gaussian Model} \cite{Cheng2012} and personalized \textit{Kernel Density Estimation} \cite{Zhang2015c}. We have also utilized geographical influence jointly with social and multi-aspect temporal factors. Similar to our prior work \cite{Hosseini2016}, we have employed the Normal Equation to minimize the error function and exploit optimized parameters of the distribution function.\\
\textbf{\textit{Temporal influence}} The time factor can be employed to promote the effectiveness of the location recommendation task either in general or the time-aware manner (proposing a new POI to the user at a specific query time). In fact, time has numerous attributes such as recency, periodicity, consecutiveness, and non-uniformness(\cite{Cheng2013a,Cho2011,Gao2013,Yuan2013,Zhao2016,Fang2016}). Based on Recency, the recommendation task \cite{Li2015} gives higher priority to the newly visited locations. Similarly, \cite{Ricci2007,Xiang2010} outline that some locations are visited steadily where others are visited merely for a short period of time (known as long-term/short-term property). Periodicity (\cite{Cho2011,Rahimi2013,Zhang2015a}) denotes that people have cyclic mobility patterns (e.g. daily return trips between home and work). 
Consecutiveness property (\cite{Zhang2015b,Cheng2013a}) states that some venues are visited sequentially (e.g. people go to the bar after the dinner). Moreover, non-uniformness declares that the check-in behavior drifts continuously during various periods (e.g. People work and amuse during weekdays and weekend respectively) (\cite{Gao2013}). From another perspective, the time factor includes a set of \textit{granular slots} (e.g. minutes, quarters, hours, days and etc.) while some are the subset of the others. Hence, we bring another aspect of the time to the recommendation which is called \textit{Temporal Subset Property} (TSP).\\
In addition, the temporal influence can be considered either discretized or continuous. Continuous manner \cite{Zhang,Yin2015} is used owing to the fact that selecting a proper time interval is not viable \cite{Yin2015}. On the contrary, as people set their schedules (e.g appointments, meetings, and etc) in a discrete style, a growing line of research \cite{Yuan2013,Gao2013,Zhao2016,Fang2016,Yin2016,Deveaud2015} has also adopted discrete-time in location recommendation. However, many works in the prior literature \cite{Zhao2016,Yin2016,Fang2016,Zhang2015b,Deveaud2015,Gao2013,Yuan2013,Zhang} integrated merely a single or two discrete intervals to avoid complexity and overfitting issues \cite{Zhang2015b}. Some methods like \cite{Gao2013} require further configurations to make the recommendation task work under specific temporal granularity.
Prior works include Matrix Factorization (\cite{Gao2013a}), Collaborative Filtering (\cite{Yuan2013}), Graph-based (\cite{Yuan2014}), and Density estimation (\cite{Zhang}). In addition, we devise a probabilistic generative model named after \textit{Multi-aspect Time-related Influence} (MATI) that can include multiple temporal slots in location recommendation and subsequently promote the performance.\\
This work distinguishes itself from our previous work \cite{Yin2015,Wang2015,Yin2015a,Yin2013,Yin2014} in the following aspects. First, we project a user's check-in behavior into a temporal latent space which predicts future visits based on current time-aware mobility patterns.; Second, we retrieve multi-aspect temporal similarity maps which both mitigates data sparsity and represents the temporal state of the user-item dataset; Third, instead of taking into account a limited number of temporal dimensions, we leverage all time-related aspects among the query user and each of proposed locations. Disregarding the level of density, this model can promote various recommendation systems through encapsulating all temporal aspects.
\section{Conclusions}
\label{conclusions}
In this paper, inspired by the fact that the discrete-time entity comprises numerous \textit{granular slots} such as minute, hour, day and etc, we proposed a novel probabilistic model, named after \textit{Multi-aspect Time-related Influence} (MATI) which simultaneously takes multiple latent temporal parameters into consideration to improve location recommendation systems. While most of the prior works utilize merely one or two limited aspects of the time, we proposed a multivariate model. On the one hand, it demotes the sparsity involved in user-location matrices in Location-based Social Networks (LBSN) and on the other hand it employs a novel \textit{Expectation-Maximization} method to compensate incomplete data w.r.t. to each latent temporal scale. Eventually, through a generalized Bayesian model, stimulated by \textit{Temporal Subset Property} (TSP), we affirmed that our approach is applicable to various types of the recommendation models.\\
To evaluate the effectiveness of our proposed approach in POI recommendation, we conducted two series of experiments. Firstly, we applied various parameter adjustments to maximize the performance of all competitive models. Consequently, we assured the effectiveness of our proposed method, both in location recommendation and succeeding the recommendation task where it is failed via the baselines. In short, we approved supremacy of our method versus various temporal and non-temporal state-of-the-art rivals.\\
The restriction involved with our proposed MATI model is that it assumes that users' temporal behavior are stable across
their check-in history. But in reality, users show various temporal mobility patterns (e.g. during travel, holidays and etc). In our future work, we will adapt our approach to studying the dynamic multivariate temporal aspect through a correlation network among each of proposing POIs and the set of previously visited location by the query user. Moreover, in order to carry out the smoothing, we will consider the fact that each temporal slot is affected by its containing latent factor. On the other hand, during hours of a day, people's behavior is different on various days of the week.
\bibliographystyle{abbrv}
\bibliography{location_modelling}
\end{document}